\documentclass[12pt,preprint,apj]{emulateapj}
\usepackage{enumerate}
\usepackage{amsmath}
\usepackage{graphicx,subfigure}

\usepackage{multirow}
\usepackage{bigstrut}

\usepackage{tablefootnote}
\usepackage[flushleft]{threeparttable}

\usepackage{verbatim} % for \begin{comment}     \end{comment}
\usepackage{color} % added
\usepackage[usenames,dvipsnames]{xcolor}

\usepackage[usenames,dvipsnames]{xcolor}
\definecolor{green}{rgb}{0.0, 0.4, 0.0}
\definecolor{forestgreen(web)}{rgb}{0.13, 0.55, 0.13}

\shorttitle{Evolution of the spin of late-type galaxies} 
\shortauthors{Hwang et al.}

\begin{document}

\title{Evolution of the spin of 
late-type galaxies caused by galaxy-galaxy interactions}

\author{Jeong-Sun Hwang$^{1,2}$, Changbom Park$^{3}$, 
Soo-hyeon Nam$^{4}$, and Haeun Chung$^{5}$}

\affil{$^1$ 
Department of Physics and Astronomy, Sejong University, Seoul 05006, Korea; hwang2k@gmail.com}
\affil{$^2$ 
Department of Science Education, Gwangju National University of Education, Gwangju 61204, Korea}
\affil{$^3$ 
School of Physics, Korea Institute for Advanced Study, Seoul 02455, Korea}
\affil{$^4$ 
Department of Physics, Korea University, Seoul 02841, Korea}
\affil{$^5$ 
University of Arizona, Steward Observatory, 933 N Cherry Ave, Tucson, AZ 85721, USA}

\begin{abstract}
We use N-body/hydrodynamic simulations to study the evolution 
of the spin of a Milky Way-like galaxy through interactions.
We perform a controlled experiment of co-planner 
galaxy-galaxy encounters and study the evolution of disk spins 
of interacting galaxies. 
Specifically, we consider the cases 
where the late-type target galaxy 
encounters an equally massive companion 
galaxy, which has either a late or an early-type morphology,  
with the closest approach distance of about 50~kpc,  
in prograde or retrograde sense. 
By examining the time change of the circular velocity of 
the disk material of the target galaxy from each case, 
we find that the target galaxy tends to lose the spin through 
prograde collisions but hardly through retrograde collisions, 
regardless of the companion galaxy type. 
The decrease of the spin results mainly from the deflection 
of the orbit of the disk material by tidal disruption.  
Although there is some disk material which gains the circular velocity  
through hydrodynamic as well as gravitational interactions 
or by transferring material from the companion galaxy, 
it turns out that the amount of the material is generally 
insufficient to increase the overall galactic spin 
under the conditions we set. 
It is found that the spin angular momentum of the target galaxy disk  
decreases by 15 - 20\% after a prograde collision. 
We conclude that the accumulated effects of galaxy-galaxy interactions 
will play an important role in determining the total angular momentum 
of late-type galaxies at current stage. 
\\

\end{abstract}

\keywords{
galaxies: spiral 
--- galaxies: kinematics and dynamics 
--- galaxies: evolution 
--- galaxies: interactions  
--- methods: numerical 
}

%====================================
%            Sec. 1 
%====================================

\section{INTRODUCTION}

The spin of galaxies is an important quantity to understand 
the evolution of galaxies over cosmic time. 
Starting from the tidal torque 
theory \citep{Hoyle1949, Peebles1969, White1984}, 
it has been proposed that galaxies acquire angular momentum 
via the tidal field from surrounding galaxies. 
In the Lambda Cold Dark Matter ($\rm\Lambda CDM$) paradigm, 
galaxies are first formed in dark matter (DM) halo  
and evolved through 
continuous merging and interacting with nearby galaxies 
while gaining or losing mass and angular momentum. 
In particular, it is vital to understand how the angular 
momentum of late-type galaxies (LTGs) are being evolved over 
cosmic time, because the LTGs are considered 
as building blocks under the hierarchical merging scenario. 
For example, the existence of a relatively new population of 
rotation-supported early-type galaxies (ETGs) can not be 
fully explained without understanding the role of LTG spin 
evolution \citep{Emsellem+2011, Cappellari2016, Graham+2018}.

Previous studies have shown that the spin of galaxies   
is also correlated with environments such as
distance to the neighbour, axes of large-scale structure, 
2D correlation, and clustering 
\citep{Porciani+2002, Davis_Natarajan2009, Tempel+2013, 
{Casuso_Beckman2015}, Codis+2015, JLee+2018}. 
More recent studies have demonstrated how the 
angular momentum acquired through merging can affect  
the galaxy evolution 
\citep{Brook+2012, Cloet-Osselaer+2014, Gomez+2017}.

There has also been some effort to measure any correlations   
between the spins in pairs of spiral galaxies. 
For example, \citet{Sodi+2010} have estimated 
the spin parameter for 3624 LTGs 
(target galaxies) at low redshift, 
identified using the Sloan
Digital Sky Survey (SDSS).   
They have discovered a statistically significant correlation 
between the spin magnitudes of neighbouring galaxies.   
Specifically, they have found that the values of 
the spin parameter of the late-type target galaxies decrease as 
a companion galaxy (the nearest neighbour of a comparable size) 
approaches, regardless of the companion morphology,  
although the decrease of the spin parameter 
appeared more clearly when the companion is a late-type  
rather than an early-type (refer to their Figures 3 and 4). 
Particularly noteworthy is that the decrease of the spin parameter  
begins to operate as soon as the (projected) separation 
between the pair becomes smaller than the virial radius 
of the companion, implying interactions could have caused the decrease.

More recently, 
using the data of Mapping Nearby Galaxies at APO (MaNGA), 
\citet{Lee+2018} have estimated the spin parameter for  
1830 low-redshift galaxies 
from the analysis of two-dimensional stellar spectra. 
They divided the galaxies into four groups by the 
morphological type of the target as well as the   
companion (the nearest influential neighbour).
As shown in their Figure~7, 
in the case of late-type targets 
having an early-type companion (``L-e"), 
the spin parameter of the target galaxies  
has been found to be decreased clearly  %as  
when the pair separation becomes less than 
the virial radius of the companion.   
(In the case of late-type targets 
having a late-type companion, ``L-l", 
such decrease has not been found clearly  
in the spin of the targets.   
However, it should be noted that the number of close pairs 
in this case is smaller than that in L-e case.)

The aforementioned observational studies that have shown 
some interesting correlations between the 
spins in pair of galaxies 
(\citealt{Sodi+2010}; \citealt{Lee+2018}) 
motivated us to explore the evolution of 
the spin of galaxies by using numerical simulations.  
With the aim of finding physical mechanism 
for changing the angular momentum of galaxies, 
we have been performing a series 
of $N$-body/hydrodynamic simulations 
of interactions between galaxies in a variety of environments. 
(We generally refer to ``angular momentum" as 
``spin angular momentum", if not specified.)

In this paper, we present the results of our first set of 
simulations concerning the evolution of 
the spin of a Milky Way-like LTG (target galaxy) 
that experiences an interaction with a 
neighbour (companion galaxy).
Specifically, we have conducted simulations 
of co-planner galaxy-galaxy interactions,   
where the target galaxy encounters   
an equally massive companion galaxy, 
which has either a late- or an early-type morphology, 
in prograde or retrograde sense.  
We also have performed a simulations of the target galaxy 
evolving in isolation for comparison. 
The simple configuration for this work is intended   
to focus on examining the effects that 
the different companion morphology (late-type vs. early-type) 
and the encounter geometry (prograde vs. retrograde) 
might have on the spin evolution of the target galaxy, 
in comparison with that of the isolated target galaxy.   
As motivated by observations, this work aims at examining 
what happened on the spin of the target galaxy.   
We expect the results of this first set of simulations 
to serve as a basis for our more extended  
follow-up studies and also for other related works.

The outline of this paper is as follows.
We begin in Section~2 with an overview of the  
galaxy models and the simulation code used for this work.  
We then describe the initial conditions of our simulations.   
In Section~3, we present the results of our simulations 
focusing on the structural and the kinematic evolution 
of the target galaxy from each of our runs. 
Finally in Section~4, we summarise and discuss our findings.  
We also highlight the limitations and the implications 
of our results.

%====================================
%            Sec. 2
%====================================

\section{Model Description}

\subsection{Galaxy Models}

For this numerical study, we use two galaxy 
models - a LTG model~``L" and an ETG model~``EH". 
Model~L is the same model used in our previous 
work (\citealt{Hwang+2018}; hereafter Paper I). 
Model~EH is similar to model~EH in paper~I but 
has half the mass of the previous model. 
Thus models~L and EH in this work 
have the same mass.  
Both models~L and EH are generated by using the 
ZENO\footnote{http://www.ifa.hawaii.edu/$\sim$barnes/software.html} 
software package (\citealt{Barnes2011}; 
see also \citealt{Hwang+2013} for the generation procedure).  
As the detailed description of the models is presented 
in Paper~I (see also \citealt{Hwang+2013}),  
here we briefly outline their key properties.

%----------<< Table 1: Model parameters >>----------
\begin{table*}[ht]
\begin{center}
\caption{Model parameters}
\label{table:tab01}
\begin{threeparttable}
\centering
\renewcommand\arraystretch{1.5} 
\begin{tabular}{c||rr|rrrrr|rr|rrrrr}
\hline
Model & 
$M_{\rm{tot}}^{a}$  & $R_{\rm{vir}}^{b}$ & 
$M_{\rm{ds}}$  & $M_{\rm{dg}}$ &  $M_{\rm{b}}$  & 
$M_{\rm{hd}}$  & $M_{\rm{hg}}$ & 
$f_{\rm{dg}}$  & $f_{\rm{hg}}$ & 
$N_{\rm{ds}}$  & $N_{\rm{dg}}$ &  $N_{\rm{b}}$  & 
$N_{\rm{hd}}$  & $N_{\rm{hg}}$ 
\\
\hline
L      & 
127.0  & 214.0    & 
5.2 & 0.8 & 1 &
120.0 & ... &
0.13 & ... & 
122 880 & 32 768 & 24 567 & 
655 360 & ...
\\
EH    & 
127.0  & 250.0    & 
... & ... & 7.0 &
118.8 & 1.2 & 
... & 0.01 & 
... & ... & 172 032 & 
655 360 & 49 152
\\
\hline
\end{tabular}
\begin{tablenotes}
\item[] %My Note.
Masses are in units of $10^{10}\,{M_{\odot}}$ 
and the virial radius is in units of kpc.  
\item[a]
$M_{\rm{tot}}$ is the sum of the masses of all particles. 
\item[b] 
We use the virial radius $R_{\rm{vir}}$ as $R_{\rm{200}}$, 
which is defined as the radius within
which the average density is 200 times the critical density.  
\end{tablenotes}
\end{threeparttable}
\end{center}
\end{table*}

\begin{figure} [!hbt]
\centering
\includegraphics
[width=6.5cm] 
{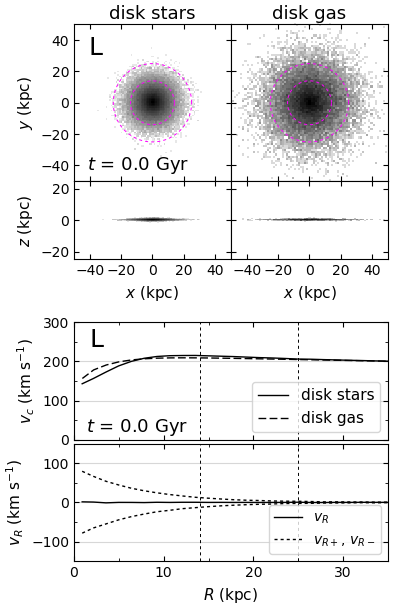} 
\caption{
Top two rows: 
initial distribution of the disk particles of model~L.  
The disk star and gas particles are
displayed separately in the left and right columns, respectively. 
The upper and lower rows show the distribution 
projected on to the x-y and x-z plane, respectively. 
The gray scale represents the column densities of the star 
and gas disks. (The gray scale varies). 
The inner and outer dotted circles (magenta) drawn 
in the upper panels indicate $R_{14}$ and $R_{25}$, respectively.   
Bottom two rows: 
initial profiles of the circular and the radial velocities 
($v_{\rm{c}}$ and $v_{\rm{R}}$, respectively) of the disk 
particles of model~L. 
In the upper panel, the solid and dashed curves represent the  
mass weighted, cylindrically averaged azimuthal velocities 
of the disk stars and gas particles, respectively,    
with respect to the cylindrical radius~$R$.  
In the rower panel, the solid curve shows the mass weighted, 
cylindrically averaged radial velocities of the stellar particles; 
the upper and lower dotted curves represent the average values of 
$v_{\rm{R+}}$ and $v_{\rm{R-}}$ of the star particles, respectively. 
For the gas, the radial velocity profile is not shown 
as it is zero at the initial time.
In each panel, the two vertical dotted lines (black) 
indicate $R_{14}$ and $R_{25}$. 
(The vertical velocity profile of the disk material 
is not shown, as it is not significant in our work.) 
} 
\label{fig:fig01}    
\end{figure}

Model~L resembles the Milky Way Galaxy in size and mass, 
with the total mass 
of $M_{\rm{tot}}$= $1.27 \times 10^{12}{M_{\odot}}$ 
and the virial radius of $R_{\rm{vir}}$ = 214~kpc 
(Table~\ref{table:tab01}). 
The model consists of four components - a stellar disk, 
a gas disk, a stellar bulge, and a DM halo. 
The star and gas disks have the masses of  
$M_{\rm{ds}}$ = $5.2 \times 10^{10}\,{M_{\odot}}$ and  
$M_{\rm{dg}}$ = $0.8 \times 10^{10}\,{M_{\odot}}$, 
respectively (\citealt{McMillan2011}; \citealt{Kubryk+2015}).   
The disk gas fraction in mass,  
$f_{\rm{dg}}$ = $M_{\rm{dg}}$/$(M_{\rm{ds}} + M_{\rm{dg}})$, 
is thus about 0.13. 
Both disks follow an exponential surface density  
profile and a sech$^2$ vertical 
profile (Equation~(1) in paper~I). 
The radial scale lengths of the stellar and gaseous 
disks are $r_{\rm{ds}}$ = 3.5~kpc and 
$r_{\rm{dg}}$ = 8.75~kpc, respectively,   
and the vertical scale length of both disks is 
$z_{\rm{ds}}$ = $z_{\rm{dg}}$ = 0.35~kpc (refer to 
\citealt{Hwang+2013} for the justifications of 
the parameter values). 
The number of particles 
distributed on the star and gas disks  
are $N_{\rm{ds}}$ = 122 880 and 
$N_{\rm{dg}}$ = 32 768, respectively. 
Both disks are initialized with 
an either clockwise or counterclockwise  
directional spin, as necessary.    
The disk gas particles are set to rotate 
with the local circular velocities. 
On the other hand, the disk star particles are given to have  
the radial and vertical dispersions in addition to the local 
circular velocities as described in \citet{Barnes_Hibbard2009}.  
In Figure~\ref{fig:fig01},  
we show the initial distribution of 
the disk star and gas particles of model~L  and 
the circular and radial velocity 
($v_{\rm{c}}$ and $v_{\rm{R}}$, respectively)  
profiles of the disk star and gas particles. 
(Hereafter, we refer to the azimuthal and the radial velocities  
in the cylindrical coordinate system  
as the circular and the radial velocities 
($v_{\rm{c}}$ and $v_{\rm{R}}$), respectively.  
The positive directions of $v_{\rm{c}}$ and $v_{\rm{R}}$ 
are set to the direction of the initial disk rotation  
and the radially outward direction, respectively.)  
The temperature of the disk gas particles are set 
uniformly to T = 10 000~K.
The bulge component has the total mass of 
$M_{\rm{b}}$ = $1.0 \times 10^{10}\,{M_{\odot}}$. 
It follows the Hernquist profile (\citealt{Hernquist1990}) with 
truncation at large radii as presented in Paper~I (Equation~(2)). 
The number of particles of the bulge component   
is $N_{\rm{b}}$ = 24 567. 
The DM halo component follows a \cite{Navarro+1996} model 
with an exponential taper at large radii (Equation~(3) in Paper~I).   
The total mass of the DM halo 
is $M_{\rm{hd}}$ = $120~\times 10^{10}\,{M_{\odot}}$  
and the number of particles is $N_{\rm{hd}}$ = 655 360.
The DM halo and the bulge components are 
dispersion-supported with no appreciable rotation 
(\citealt{Barnes_Hibbard2009}; \citealt{vonNeumann1951}).

The ETG model~EH has the total mass of 
$M_{\rm{tot}}$=$1.27 \times 10^{12}{M_{\odot}}$   
(which is equal to that of model~L, as intended) 
and the virial radius of $R_{\rm{vir}}$ = 250~kpc. 
The model possesses three components, a stellar bulge, 
a DM halo, and a gaseous halo, with the masses of    
$M_{\rm{b}}$ = $7.0 \times 10^{10}\,{M_{\odot}}$,    
$M_{\rm{hd}}$ = $118.8 \times 10^{10}\,{M_{\odot}}$, and 
$M_{\rm{hg}}$ = $1.2 \times 10^{10}\,{M_{\odot}}$ 
(\citealt{Anderson_Bregman2010}), respectively.   
The gas fraction in the halo is 
$f_{\rm{hg}}$ = $M_{\rm{hg}}$/$(M_{\rm{hd}} + M_{\rm{hg}})$ = 0.01.  
The bulge and the DM halo components follow 
the Hernquist model (\citealt{Hernquist1990}) and 
a \cite{Navarro+1996} model, respectively, 
and are dispersion-supported as in model~L. 
The gas halo follows a nonsingular isothermal profile 
with a taper (Equation~(3) in paper~I). 
The temperature of the halo gas particles are determined 
by the hydrostatic equilibrium (refer to Figure~2 of Paper~I). 
The initial velocity of the halo gas is set to zero.    
The number of particles distributed in each component 
is $N_{\rm{b}}$ = 172 032, $N_{\rm{hd}}$ = 655 360, 
and $N_{\rm{hg}}$ = 49 152, respectively.

%----------<< Figure 01: initial model~L >>---------- 

\subsection{Simulation Code}

To perform the numerical simulations,  
we make use of an early version of GADGET-3 
$N$-body/smoothed particle hydrodynamics (SPH) code 
(originally described in \citealt{Springel2005}). 
Here we briefly describe the simulation code and refer
interested readers to \citet{Springel_Hernquist2003} and 
our previous papers for more details 
(paper~I; \citealt{Hwang_Park2015}).

The code computes the gravitational force using 
a tree algorithm (\citealt{Barnes_Hut1986}) 
and the hydrodynamic force adopting a SPH method 
in the entropy conservative formulation 
(\citealt{Springel_Hernquist2002}; \citealt{Gingold_Monaghan1977}).  
The radiative cooling and heating are modeled 
considering the primordial mixture of hydrogen and helium 
by photoionization (\citealt{Katz+1996}).
Star formation and the associated supernova feedback 
in the interstellar medium (ISM) are implemented 
adopting the sub-resolution multiphase model of  
\citet{Springel_Hernquist2003}.
A thermal instability operates for gas exceeding a threshold 
density and the ISM is treated as a statistical mixture of 
cold clouds and ambient hot medium.
Stars form in dense regions consuming the cold
clouds, and the consumption timescale is chosen to match the
observations (\citealt{Kennicutt1998}). 
Among the newly formed stars, 
massive stars (with the mass greater than 8~${M_{\odot}}$) die 
instantly as supernovae and release energy as 
heat to the ambient diffuse gas. 
Some cold clouds are also evaporated inside the supernova bubbles
returning material to the ambient medium.

We set the code parameters governing star formation 
and feedback to the standard values of the 
multiphase model (\citealt{Springel_Hernquist2003}). 
Some of the key values are as follows.  
The star formation time-scale and the mass fraction of massive stars 
among the newly formed stars are   
$t_{0}^{\star}$ = 2.14~Gyr and $\beta = 0.1$, respectively. 
The ``supernova temperature" is  $T_{\rm SN} = 10^{8}$ and 
the temperature of cold clouds are $T_{\rm c} = 1000$~K.  
The parameter value for supernova evaporation is $A_{0} = 1000$.

The gravitational softening lengths for the particles of 
the star and gas disks, bulge, and DM and gas halos are  
set to 0.14, 0.11, 0.14, 0.30, and 0.11~kpc, respectively. 
These gravitational softening lengths are determined 
by considering the equivalence of the maximum 
acceleration experienced by 
a single particle in each component.

%----------<< Table 2: runs >>----------
\begin{table}[ht]
\begin{center}
\caption{Simulations}
\label{table:tab02}
\begin{threeparttable}
\centering
\renewcommand\arraystretch{1.5} 
\begin{tabular}{l|lll}
\hline
Run & Target (spin\tnote{a} ) 
    & Companion (spin) 
    & Interaction\tnote{b}        \\
\hline
L      & L (CW)  & ...    & ...  \\
LLp    & L (CW)  & L (CW) & prograde\\
LLr    & L (CCW) & L (CW) & retrograde\\
LEHp    & L (CW)  & EH     & prograde\\
LEHr    & L (CCW) & EH     & retrograde\\
\hline 
\hline
\end{tabular}
\begin{tablenotes}
\item[a] The direction of the spin of the disk of a LTG, 
when viewed from the positive $z$-axis. 
``CW" and ``CCW" stand for clockwise and 
counterclockwise, respectively. 
\item[b] Whether a target galaxy experiences a collision 
with a companion galaxy 
as prograde or retrograde (see text for the details).  
\end{tablenotes}
\end{threeparttable}
\end{center}
\end{table}

%----------<< Figure 2: orbital trajectories >>----------
\begin{figure} [!hbt]
\centering
\includegraphics[width=8.5cm] 
{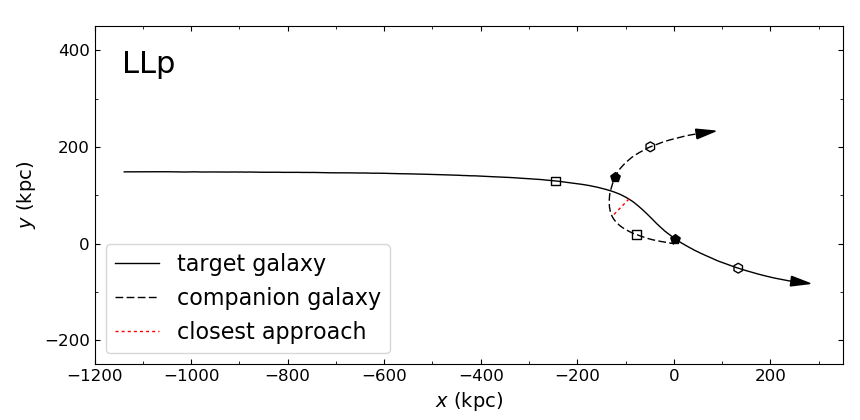}
\caption{
Orbital trajectories of the target (black solid line) 
and companion (black dashed line) galaxies  
in run~LLp during 7~Gyr. 
(The orbital trajectories from the other runs~LLr, 
LEHp, and LEHr are 
only slightly different from those shown here.) 
The target galaxy starts from  
($x_{0},y_{0},z_{0})=(-1140~\rm{kpc}, 150~\rm{kpc}, 0)$ 
with the initial velocity of 
($v_{x0},v_{y0},v_{z0}$)=$(200~\rm{km}~s^{-1}, 0, 0)$ 
and the companion galaxy from the origin 
with zero initial velocity. 
The target and companion galaxies meet most closely 
at $t =$ 4.5~Gyr with the distance of   
about 50~kpc (red dotted line).   
The positions of the two galaxies at $t =$ 4, 5, and 6~Gyr   
are marked on their orbits with open squares, filled pentagons, 
and open hexagons, respectively. 
}
\label{fig:fig02} 
\end{figure}

\subsection{Initial conditions for our simulations}

In order to study the spin evolution of a Milky Way-like 
LTG in different situations, 
we perform five numerical simulations 
(Table~\ref{table:tab02})\footnote{Additionally, we have 
performed ``run~LLr2".   
In this run, the late-type target galaxy encounters 
the late-type companion galaxy in retrograde sense, 
while both galaxies spin in counterclockwise direction. 
We present the results of this run in Appendix~A.}: 
Run~L is the simplest one where the LTG model~L (target galaxy) 
evolves in isolation. 
In runs LLp and LLr, model~L (target) encounters   
another model~L (companion) in prograde and 
retrograde sense, respectively. 
Similarly, in runs LEHp and LEHr, model~L (target) 
encounters the ETG model~EH (companion) in prograde and 
retrograde sense, respectively.   
We construct the initial conditions of our simulations as below.

First, for the isolated case of run~L, 
model~L is initialized in the $x$-$y$ plane 
(i.e., the midplane of the disk lies on the $x$-$y$ plane), 
being placed the center of the model at the origin 
as shown in Figure~\ref{fig:fig01} (top two rows).  
(We generally refer to ``disk" as the total disk 
consisting of both stars and gas, if not specified 
such as ``star disk" or ``gas disk".) 
The disk of model~L in this run is set 
with a clockwise directional spin 
as viewed from the positive $z$-axis, 
and with zero (systemic) velocity.

Next, for runs~LLp and LLr, the two LTG-LTG encounter simulations,    
model~L as the target galaxy is initially placed at 
($x_{0},y_{0},z_{0}$)=($-$1140~kpc, 150~kpc, 0)   
and another model~L as the companion galaxy is positioned at 
the origin as presented in Figure~\ref{fig:fig02}.   
The initial velocity ($v_{x0},v_{y0},v_{z0}$) of the target 
and the companion galaxies are set 
to $(200~\rm{km}~s^{-1}, 0, 0)$ and  $(0, 0, 0)$, respectively. 
These initial positions and velocities of the models 
are chosen so that the two galaxies encounter most 
closely at $t$ = 4.5~Gyr since the start of each run  
with the distance of about 50~kpc. 
The closest approach distance of 50~kpc is chosen 
to see some encounters that are not too weak or not too strong. 
For galaxies at low redshift and not in high-density regions 
like galaxy clusters, which are relevant to this study,    
distant encounters occur more frequently 
than very close encounters do. 
Although distant encounters are frequent, 
the effects of a single distant encounter on the evolution 
of a target galaxy would be weak. 
On the other hand, the effects of a single very close encounter 
would be significant, but such an encounter is rare.  
(Once it happens, the galaxies involved would merge quickly.) 
Thus, considering both frequency and effects of 
galaxy-galaxy interactions, choosing encounters at 
an ``intermediate" strength would be reasonable for our work. 
In the initial conditions of runs~LLp and LLr, 
the midplanes of the disks of both target and companion galaxies 
are located on the $x$$-$$y$~plane. 
The only difference between the two initial conditions  
is the direction of the spin of the target galaxy:  
For run~LLp, the disk of the target galaxy is set with  
a clockwise directional spin; whereas, for run~LLr, 
it is set with a counterclockwise directional spin. 
In both runs, the disk of the companion galaxy is  
set with a clockwise directional spin.

Finally, for runs~LEHp and LEHr, the two LTG-ETG encounter runs,  
the initial positions and velocities of the target and the companion 
galaxies are set with the same values as in runs~LLp and LLr.
The direction of the spin of the target galaxy in run~LEHp(LEHr)  
is also set to clockwise(counterclockwise), as in run~LLp(LLr). 
The only difference between the initial conditions 
of run~LEHp(LEHr) and run~LLp(LLr) 
is the morphological type of the companion galaxy, 
which is an early-type in runs~LEHp and LEHr 
but a late-type in runs~LLp and LLr.

%----------<< Figure 3: ss_L >>----------

\begin{figure*} [!hbt]
\centering
\includegraphics
[width=18.0cm]
{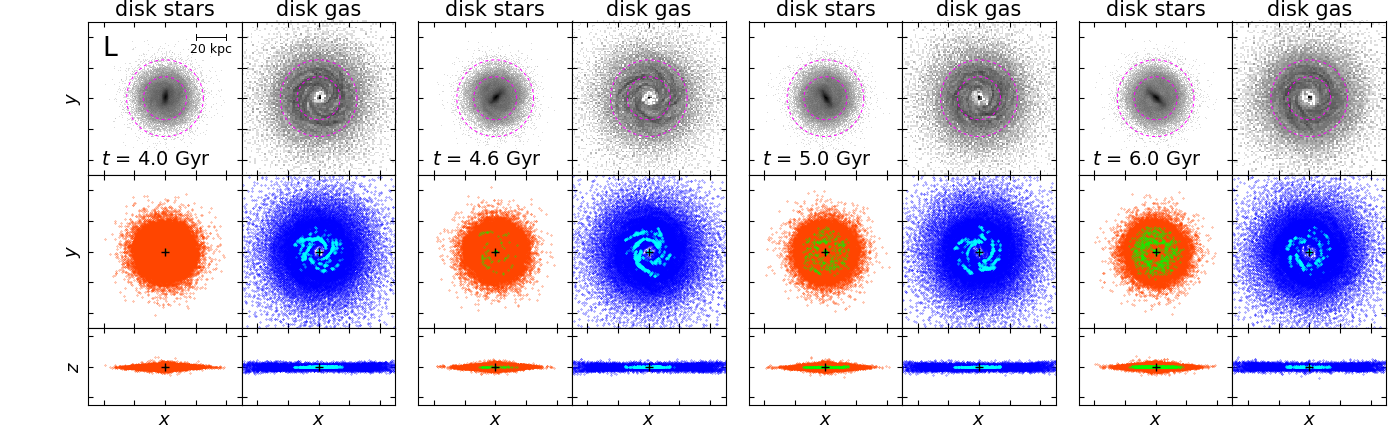}
\caption{
Four snapshots from run~L of the disk particles of the 
target galaxy at $t$ = 4, 4.6, 5, and 6~Gyr 
(from leftmost two columns to rightmost two columns 
in order, respectively; refer to Figure~\ref{fig:fig01} 
for the initial particle distribution).   
The star and gas particles 
at each epoch are displayed separately 
(left and right columns, respectively).    
The top row shows the column densities of the disks  
projected on to the $x$-$y$ plane, 
with adaptive gray scales.   
The inner and outer dotted (magenta) circles  
drawn in each panel 
of the top row indicate 
$R_{14}$ and $R_{25}$, respectively.  
The middle and bottom rows display   
the individual disk particles  
seen in the $x$-$y$ and $x$-$z$ planes, respectively, 
indicating the ``status" of the particles 
with different colors: 
Orange points represent ``old" disk stars  
(the stars originally set
as the disk stars and the stars formed out of the disk gas 
before $t =$~4.5~Gyr) and 
green points denote ``young" disk stars 
(the stars formed out of the disk gas at $t$~$\ge$ 4.5~Gyr).
Blue and cyan points represent the non-star-forming 
(gas with zero star formation rate at the instant)  
and star-forming disk gas particles (gas having positive values 
of star formation rate at the instant), respectively. 
In each panel of the middle and bottom rows, 
the center of the target galaxy   
is marked with a plus sign. 
}
\label{fig:fig03} 
\end{figure*}

%----------<< Figure 4 : SFR >>----------
 
\begin{figure*} [!hbt]
\centering\includegraphics 
[width=12cm]
{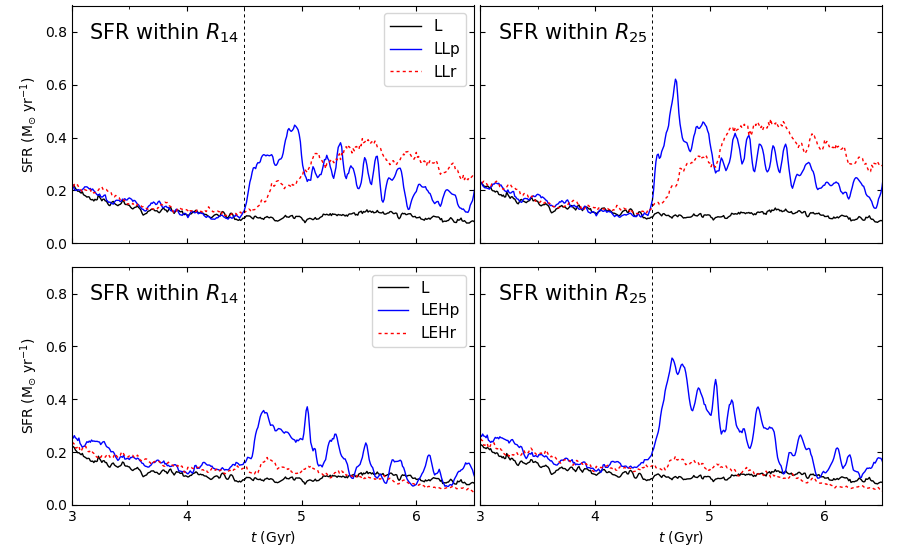}
\caption{
Evolution of the star formation rate (SFR) of the target galaxy 
from runs~L, LLp, and LLr (top row) 
and from runs~L, LEHp, and LEHr (bottom row).
The left column shows the sum of the SFRs of the 
star-forming gas particles 
enclosed in a cylindrical regions of 
radius $R_{14}$ (with $|z| \leq$ 20~kpc), 
including the captured particles from the companion  
if there are any,  
while the right column displays those enclosed in 
a cylindrical region of 
radius $R_{25}$ (with $|z| \leq$ 20~kpc) from each run. 
The vertical dotted line in each panel indicates 
the closest approach time ($t = $~4.5~Gyr) between the target 
and the companion galaxies in the galaxy-galaxy encounter runs.
}   
\label{fig:fig04} 
\end{figure*}

%----------<< Figure 5 : large-scale snapshots >>----------
 
\begin{figure*} [!hbt]
\centering\includegraphics 
[width=15cm]
{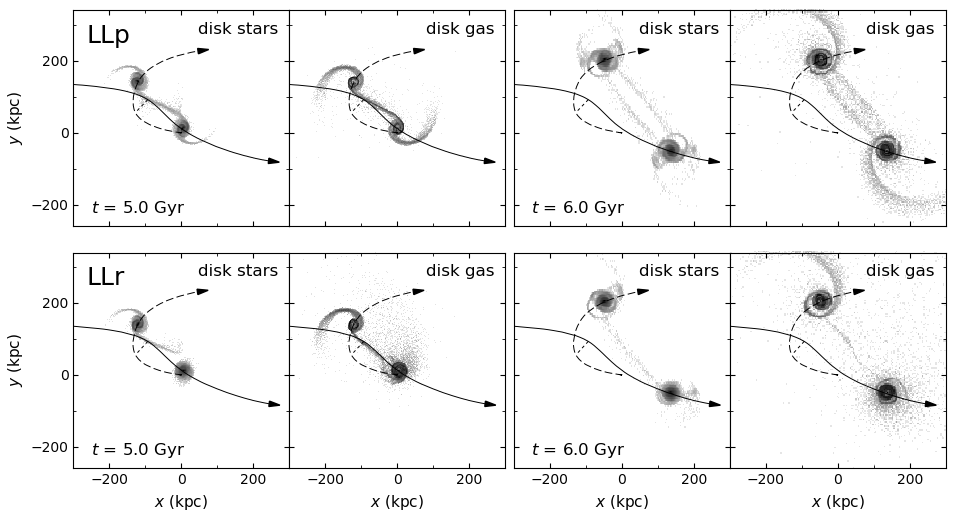}
\caption{
Large-scale snapshots from runs~LLp (top row)  
and LLr (bottom row)  
at $t$ = 5 (left two columns) and 6~Gyr (right two columns). 
In the snapshots from each run at each epoch, 
disk star and gas particles 
from both target and companion galaxies are presented   
separately (left and right panels, respectively). 
In each panel, the orbital trajectories of the target 
and companion galaxies, and the closest approach 
distance between the two galaxies 
are over-plotted (refer to Figure~\ref{fig:fig02}).  
}
\label{fig:fig05} 
\end{figure*}

%====================================
%            Sec. 3
%====================================

\section{Simulation results}

%----------<< Figure 6 : snapshots >>----------

\begin{figure*} [!hbt]
\centering
\vspace{0.2cm}
\includegraphics 
[width=18.0cm]
{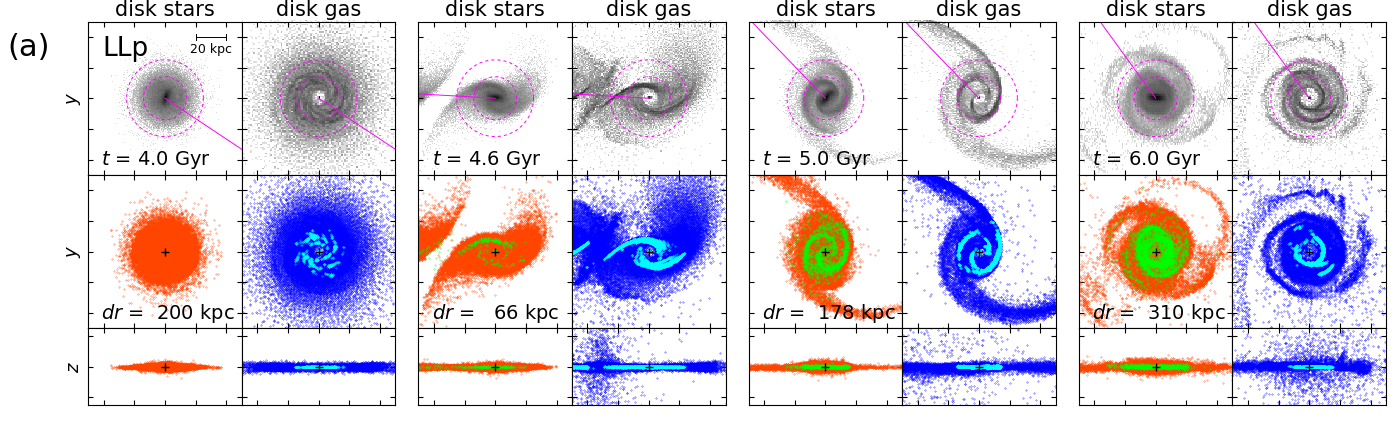}
\\ 
\includegraphics
[width=18.0cm] 
{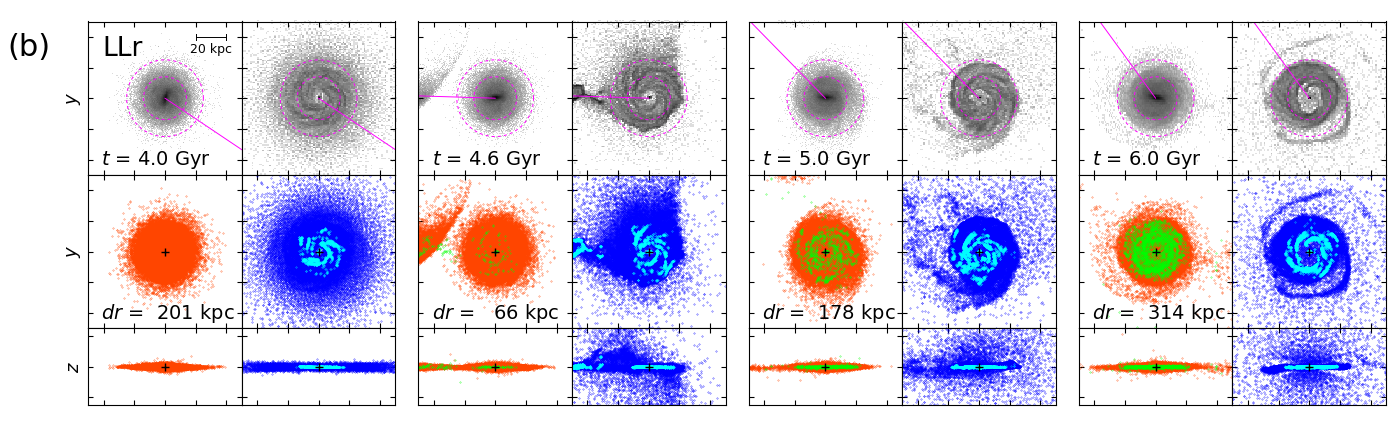}
\\ 
\includegraphics
[width=18.0cm]
{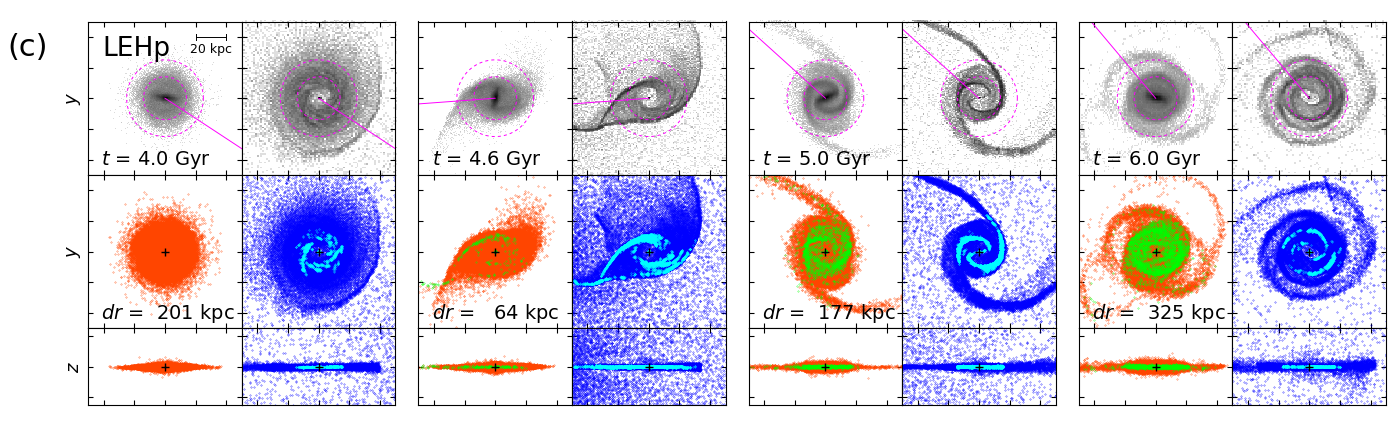}
\\ 
\includegraphics
[width=18.0cm]
{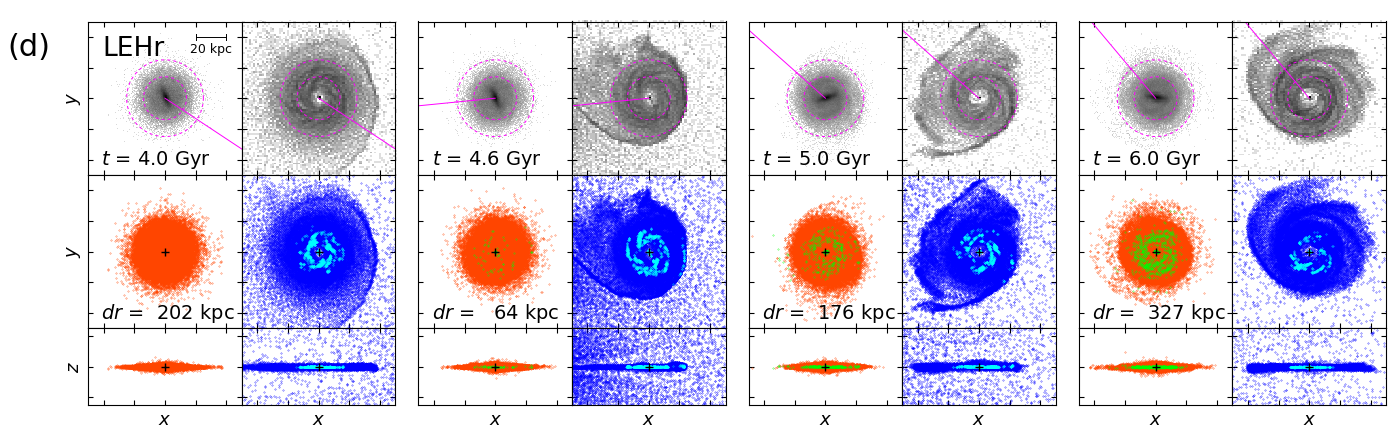}
\\
\caption{
Same as Figure~\ref{fig:fig03}, but 
for runs~LLp (a), LLr (b), LEHp (c), and LEHr (d).   
The snapshots presented here  
include any disk particles captured by the target galaxy 
from the companion galaxy, if there are any. 
In each panel of the top row of (a)-(d), the solid (magenta) line  
is drawn from the center of the target to the center of the companion.  
The distance between the two galaxies ($dr$) 
at each epoch is shown in the middle row of (a)-(d). 
}
\label{fig:fig06} 
\end{figure*}

With the initial conditions described in Section~2.3, 
we have performed   
five numerical simulations for $t =$ 7~Gyr 
since the start of each run (Table~\ref{table:tab02}). 
Here we present the results of our simulations.    
In the first subsection, we describe the general 
evolution of each run, 
focusing on the change in structure and star formation 
activity in the target galaxy.       
Then we examine the kinematics 
and the spin angular momentum of the target galaxy 
in the following subsections.

\subsection{General Evolution}

\subsubsection{Run~L}

Figure~\ref{fig:fig03} presents the distribution 
of the disk particles of the target galaxy 
from run~L at four epochs.    
The target galaxy in this run evolves in isolation. 
As seen in the top row, 
a central bar and multiple spiral patterns 
develop in the star and gas disks as time passes. 
In the bottom two rows, 
star-forming gas particles are distinguished from 
non-star-forming gas particles 
by cyan and blue points, respectively.   
Most of star-forming gas appears  
in the relatively inner region of the disk, 
within the galactocentric (cylindrical)  
radius of $R$~=~14~kpc ($R_{14}$),  
marked with a magenta dotted circle (the inner one).  
$R_{14}$ is the radius within which $\sim$90~\% 
of the disk stars were initially distributed in the model; 
it is also four times the initial radial scale length 
of the stellar disk.    
We will use $R_{14}$ as a reference radius throughout this paper, 
for the purpose of indicating the region 
where most initial disk stars were located. 
We will also use another reference radius $R$ = 25~kpc 
($R_{25}$; the outer dotted circle in the figure); 
it is half of the closest approach distance in the 
galaxy-galaxy encounter runs.

As star-forming gas subsequently turns into stars, 
the gaseous disk dissipates and the stellar disk grows in mass.  
Among those stars formed out of the star-forming gas, 
the stars formed after $t$ = 4.5~Gyr are displayed with 
green points (hereafter,``young" (disk) stars), 
whereas those stars added onto the disk before $t$ = 4.5~Gyr 
as well as the stars initially set as the disk stars 
are plotted with orange points (hereafter, ``old" (disk) stars).  
The reason we divide the young and old stars 
based on whether they were formed after $t$ = 4.5~Gyr or not 
is to compare the distribution of the young stars in  
the isolated target galaxy with 
those in the target galaxies 
from the galaxy-galaxy encounter runs 
in the following subsections. 
We present in Figure~\ref{fig:fig04} the sum of the SFRs 
of the individual star-forming gas particles 
within $R_{14}$ and $R_{25}$. 
The SFR within $R_{25}$ (and also that within $R_{14}$)  
does not greatly change,   
but generally decreases slowly as time passes, 
due to the accumulated effects of the cold gas consumption 
by star formation.

\subsubsection{Run~LLp}

Figure~\ref{fig:fig05} (top row) shows 
the large-scale configuration of the disks 
of both target and companion galaxies 
from run~LLp, 
with their orbital trajectories overlaid. 
The closest approach between the two LTGs 
occurs at $t$ = 4.5~Gyr with the distance of about 50~kpc. 
When the target galaxy collides with the companion galaxy, 
it experiences the encounter as prograde,  
since the orbit of the companion is prograde with respect to 
the clockwise directional spin of the target. 
(By the same token, the companion galaxy, 
which is spinning in clockwise direction, 
also experiences the encounter with the target as prograde.)  
Thus, as seen in the large-scale snapshots 
taken after the collision, 
substantial bridges and counter-tails are pulled out of 
the stellar and gas disks of the target(companion) galaxy 
due to the tidal force imposed by the companion(target) galaxy. 
There is some mass exchange between the galaxies:  
Some disk star and gas particles, 
among those initially set as the disk particles of the companion,  
transfer to the target through the stellar and gas bridges 
formed out of the companion, and vice versa.

Focusing on the evolution of the target galaxy,  
Figure~\ref{fig:fig06}(a) provides more close-up views 
of the star and gas disks. 
The snapshots at $t$ = 4~Gyr 
(i.e., 0.5~Gyr before the collision) appear to be similar 
to those from run~L (without a collision) at the same epoch.  
However, the snapshots taken after the collision 
show big differences compared with those from run~L  
at the corresponding epochs:   
At the onset of the collision,  
the star and gas disks of the target galaxy 
start to be pulled radially outward due to the tidal force. 
Here the tidal force not only stretches the target along the 
line connecting the target and the companion, but also compresses 
the target along the perpendicular direction. 
Besides, since gas interacts not only gravitationally 
but also hydrodynamically, 
the leading side of the gas disk of the target galaxy is 
shock-compressed by the collision with the gas disk 
of the companion galaxy. 
As seen in the snapshots 
at $t$ = 4.6~Gyr (shortly after the collision), 
a lot of star-forming gas turns up along the 
interaction-induced arms through the prograde collision.  
By $t$ = 5~Gyr, substantial bridges and tails have developed 
out of both stellar and gas disks of the target galaxy. 
The higher gas density at the leading side of 
the gas disk and of the gas bridge 
results from the shock generated through the collision. 
Until $t$ = 6~Gyr since the time of the collision, 
a lot more young stars have been added onto the disk,  
than in the isolated target galaxy from run~L. 
As shown in Figure~\ref{fig:fig04}, 
the SFR in the target galaxy increases abruptly 
right after the collision, indicating 
that the strong star formation activity 
is triggered by the prograde collision. 
The SFR within $R_{25}$ (considering the wider region) reaches 
the maximum after a few hundreds of mega years since the collision, 
and then starts to drop (but still much higher 
than that in run~L until $t$ = 6~Gyr) as gas is consumed 
by the burst of star formation.

\subsubsection{Run~LLr}

The results from run~LLr are presented in 
Figures~\ref{fig:fig05} (bottom row) and \ref{fig:fig06}(b).
When the collision occurs at $t$ = 4.5~Gyr,   
the target galaxy spinning counterclockwise direction 
experiences the collision as retrograde.  
(On the other hand, the companion galaxy spinning 
clockwise direction experiences it as prograde.) 
Due to the retrograde passage of the companion in this run    
(with respect to the spin of the target), 
tidal perturbations on the target by the companion 
act for a shorter time 
than in the case of the prograde collision (run~LLp).
Accordingly, the target galaxy here 
does not develop strong bridges and tails  
as seen in the large-scale snapshots.   
During the collision, some disk material which initially 
belonged to the companion is captured by the target galaxy.

Looking at the target galaxy more 
closely (Figure~\ref{fig:fig06}(b)),  
the amount of star-forming gas increases 
shortly after the collision (at $t = $~4.6~Gyr),  
more than in the isolated target galaxy from run~L, 
but not as greatly as in the target galaxy from run~LLp. 
It can also be checked from the SFR 
presented in Figure~\ref{fig:fig04}. 
The SFR (within $R_{25}$) of the target galaxy 
in this run~LLr increases after the collision,    
but less drastically than that in run~LLp. 
Because of the less drastic increase of the star formation activity, 
cold gas in the target galaxy 
is consumed less quickly than in run~LLp.    
This helps the target galaxy 
to maintain relatively active star formation until later times;   
the gas particles captured by the target galaxy 
during the collision also contribute to it. 
(The amount of gas captured by the target galaxy  
in run~LLr is greater than that in run~LLp, 
because in this particular case of run~LLp   
the shock boundary generated between the two gas disks 
by the prograde collision hinders the flow of gas for a while. 
In general, the amount and the orbit of the material captured by 
one galaxy from the other are very sensitive to various factors 
such as the relative orbit of the pair, the physical properties 
of each galaxy, and so on. 
A detailed discussion on this subject is beyond 
the scope of this paper.)

\subsubsection{Runs~LEHp and LEHr}

The snapshots from runs~LEHp and LEHr are presented 
in Figure~\ref{fig:fig06}(c) and (d), respectively. 
In run~LEHp(LEHr), the target galaxy spinning in 
clockwise(counterclockwise) direction experiences the collision 
with the early-type companion as prograde(retrograde).
As seen in the snapshots,  
the distribution of the old disk stars 
(which are mainly affected by the gravitational force)  
in the target galaxy from run~LEHp(LEHr) at each epoch is 
generally similar to that from run~LLp(LLr) 
at the corresponding epoch, as expected.   
However, the distributions of the disk gas 
(which is influenced by both gravitational and hydrodynamic forces)  
and of the young stars (which are formed after the collision) 
are somewhat different from those from run~LLp(LLr). 
The different hydrodynamic interactions  
between the disk gas of the target and 
either the halo gas of the companion (in runs~LEHp and LEHr)    
or the disk gas of the companion (in runs~LLp and LLr) 
may have caused the different distributions.

In run~LEHp (Figure~\ref{fig:fig06}(c)),  
at $t = 4$~Gyr (0.5~Gyr before the closest approach), 
the gas disk of the target galaxy already starts to 
form a bow-like front as it comes under the influence of 
the extended gas halo of the companion galaxy.   
At $t = 4.6$~Gyr, shortly after the closest approach, 
the gas at the leading side of the disk is shock-compressed 
by the collision (c.f. run~EH-L in \citealt{Hwang_Park2015}).  
In addition, some disk gas is stripped by ram pressure 
(as seen at the rear side of the disk) 
or ionized and then scattered,   
as it moves against the diffuse hot halo gas of the companion. 
As displayed in Figure~\ref{fig:fig04}, 
the SFR (within $R_{25})$ of the target galaxy in run~LEHp  
increases greatly by the prograde collision and then 
slows down afterwards due the cold gas consumption.

In the case of the retrograde collision 
(Figure~\ref{fig:fig06}(d)), 
the target galaxy in run~LEHr does not develop 
bridges and tails out of the star and gas disks 
after the collision.   
The SFR of the target galaxy (Figure~\ref{fig:fig04})  
increases during the collision but only slightly, 
compared with that of the target galaxy 
without a collision (run~L).   
At later times ($t \gtrsim$~5.5~Gyr), the SFR in this run  
becomes even lower than that of the isolated target galaxy. 
It is because some cold gas is removed 
not only by star formation but also 
by ram pressure and/or ionization in run~LEHr.

%----------<< Figure 7: vc >>----------

\begin{figure*} [!hbt]
\centering
\includegraphics 
[width=17.0cm]
{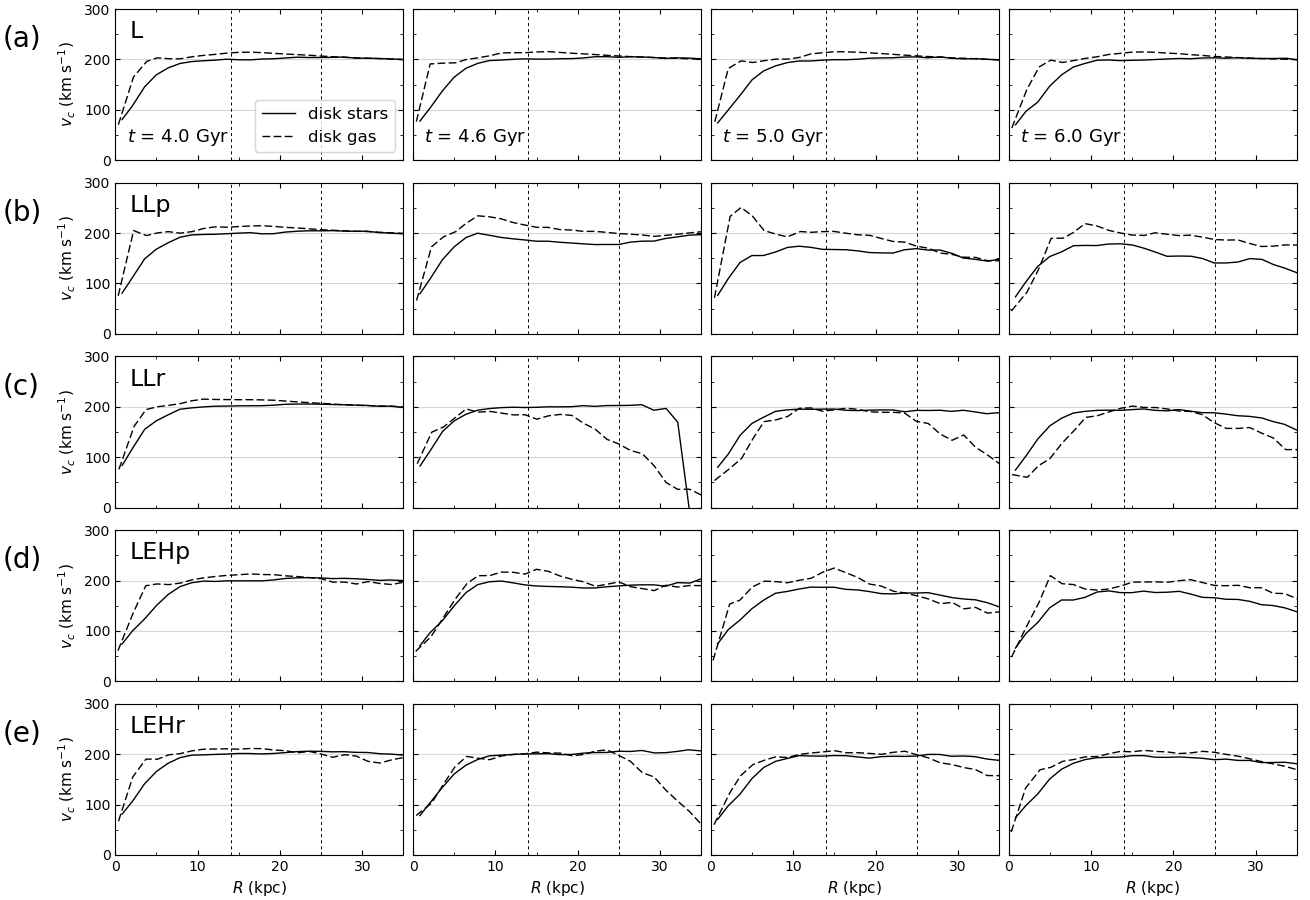}
\caption{
Mass weighted, cylindrically averaged  
circular velocity $v_{\rm c}$   
of the disk particles of the target galaxy 
with respect to the cylindrical radius $R$ 
from runs~L (a), 
LLp (b), LLr (c), LEHp (d), and LEHr (e). 
The leftmost through rightmost columns show 
the circular velocity profiles  
at $t$ = 4, 4.6, 5, and 6~Gyr, respectively.   
In each panel, solid curve represents 
the profile of the disk star particles 
(i.e., both old and young disk stars, if there are 
any young stars formed by the time) 
and dashed curve displays that of the disk gas particles 
(i.e., both non-star-forming and star-forming gas). 
The two vertical dotted lines mark $R_{14}$ and $R_{25}$. 
In (b) through (e), the velocity profiles 
are estimated including the captured particles 
(i.e., those originally set as the disk stars or gas particles 
of the companion but transferred to the target), 
if there are any.
}
\label{fig:fig07} 
\end{figure*}

%----------<< Figure 8  vR  >>----------
 
\begin{figure*} [!hbt]
\centering\includegraphics
[width=17.0cm]
{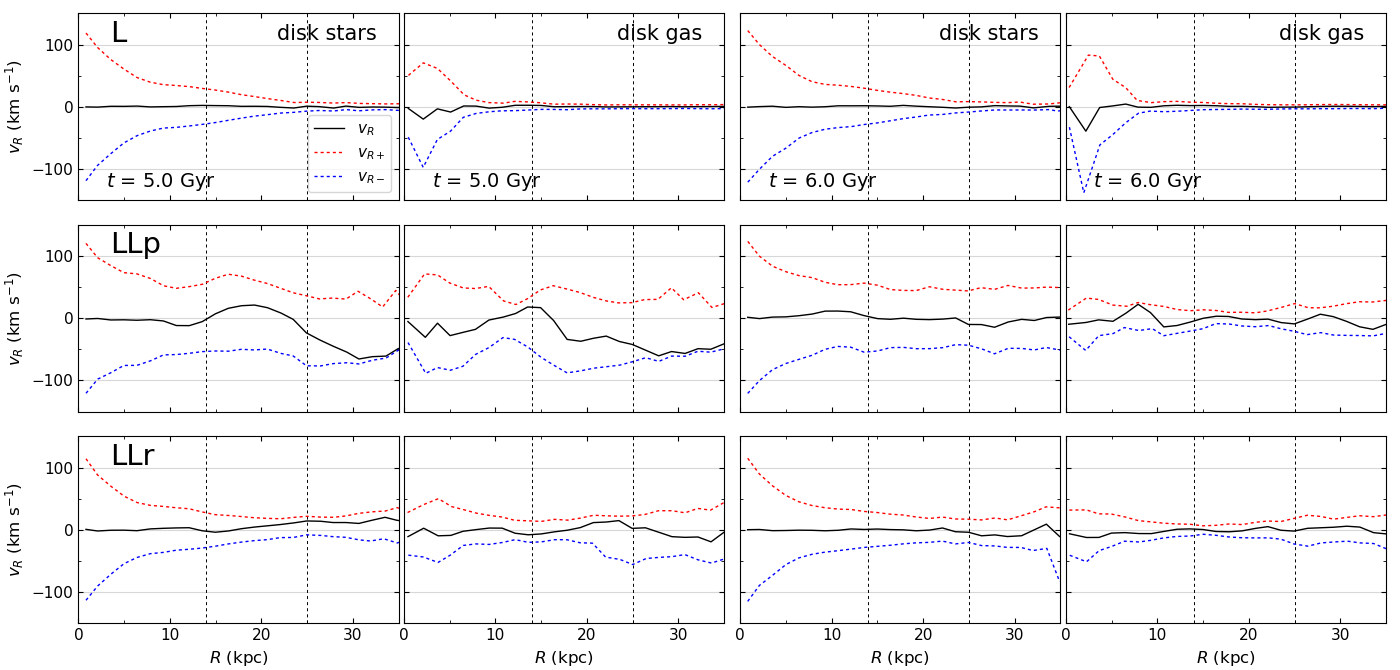}
\caption{
Mass weighted, cylindrically averaged 
radial velocity $v_{\rm R}$ 
of the disk particles of the target galaxy 
with respect to the cylindrical radius $R$ 
from runs~L (top row), 
LLp (middle row), and LLr (bottom row). 
The left two columns present  
the velocity profiles of the stars and gas, 
respectively, at $t$ = 5~Gyr; 
similarly, the right two columns show those of the 
stars and gas at $t$ = 6~Gyr. 
In each panel, 
the black solid curve represents the profile of $v_{\rm R}$;  
the red and blue dotted curves show 
those of $v_{\rm R+}$ and $v_{\rm R-}$ separately.    
The two vertical dotted lines 
indicate $R_{14}$ and $R_{25}$.
In the middle and bottom rows, 
the velocity profiles are estimated including 
the captured particles from the companion galaxy,    
if there are any. 
}
\label{fig:fig08} 
\end{figure*}

%----------<< Figure 9 vc: stars -- stars+gas >>----------
 
\begin{figure*} [!hbt]
\centering\includegraphics
[width=17.0cm]
{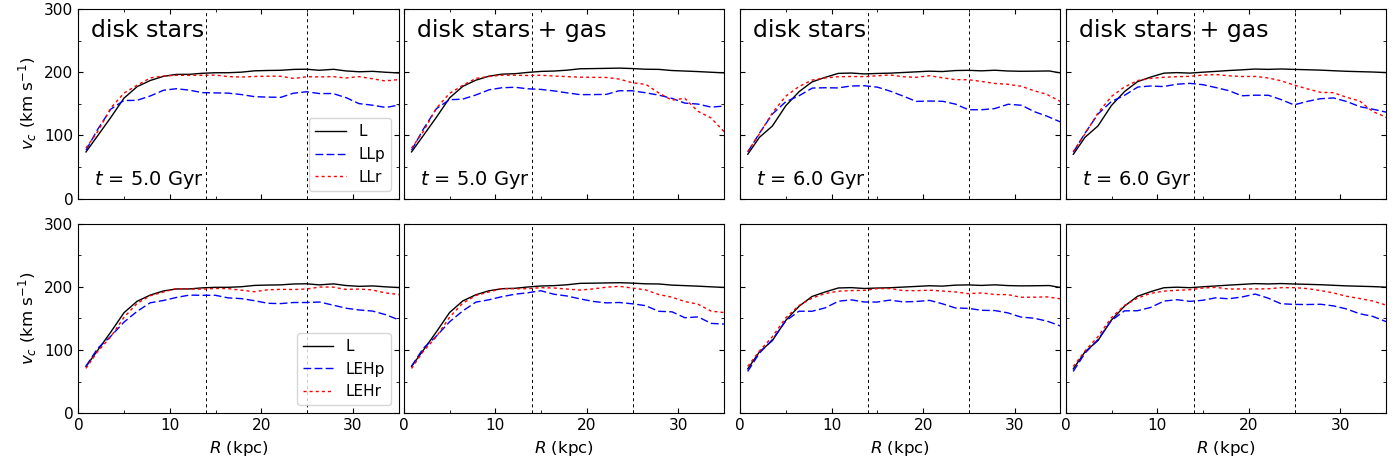}
\caption{ 
Top row: comparison of the circular velocity profiles 
of the disk material of the target galaxy from 
runs~L, LLp, and LLr (black solid, blue dashed, 
and red dotted curves, respectively).  
The left two columns show the profiles of the disk star particles  
and of the all disk (stars + gas) particles, respectively, 
at $t = $~5~Gyr; similarly, the right two columns 
represent those of the stars and of the 
star + gas, respectively, at $t = $~6~Gyr.       
In each panel, the two vertical dotted lines  
mark $R_{14}$ and $R_{25}$. 
Bottom row: same as the top row, but for runs~L, LEHp, and LEHr. 
}
\label{fig:fig09} 
\end{figure*}

%----------<< Figure 10 : vcabs particles >>----------

\begin{figure*} [!hbt]
\centering
\includegraphics
[width=17.5cm]
{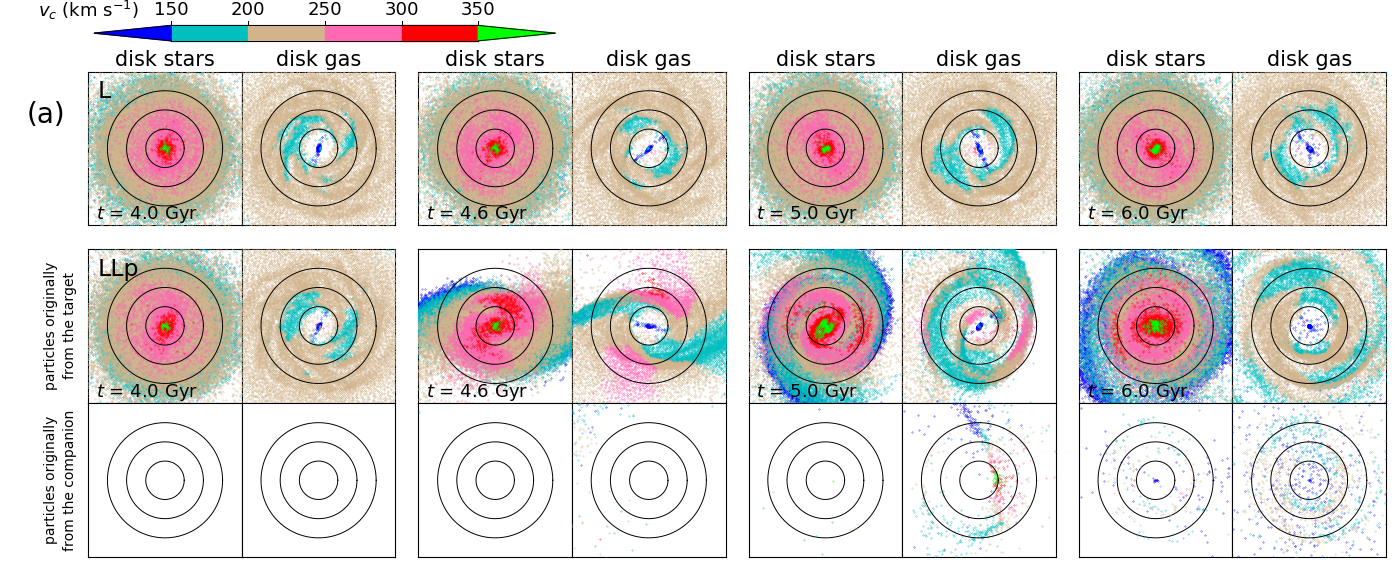}
\vspace{0.3cm}
%---------------------
\includegraphics
[width=8.5cm]
{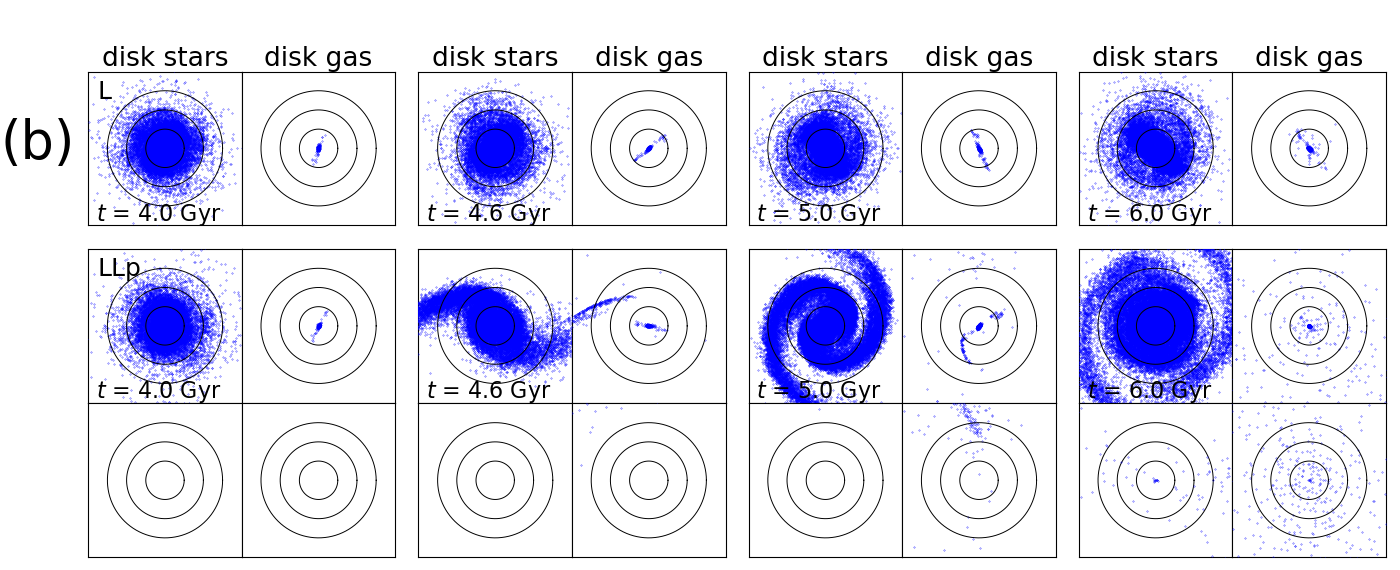}
\;\;  %\quad
\includegraphics
[width=8.5cm]
{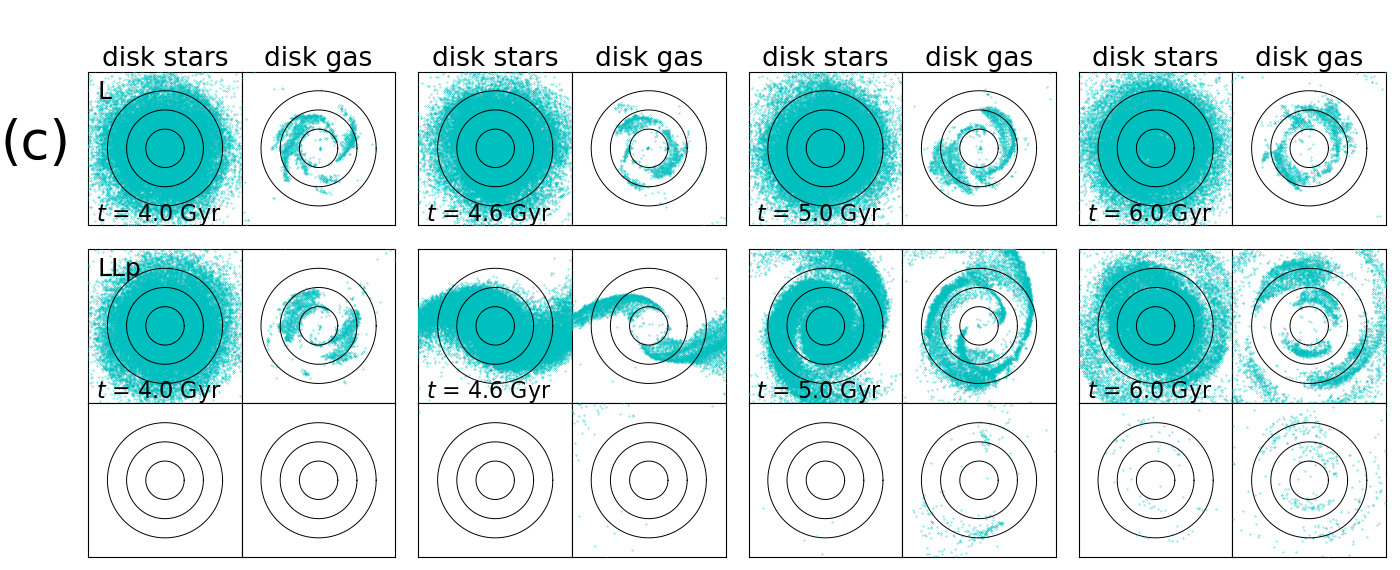}
\\
\vspace{0.3cm}
%---------------------
\includegraphics
[width=8.5cm]
{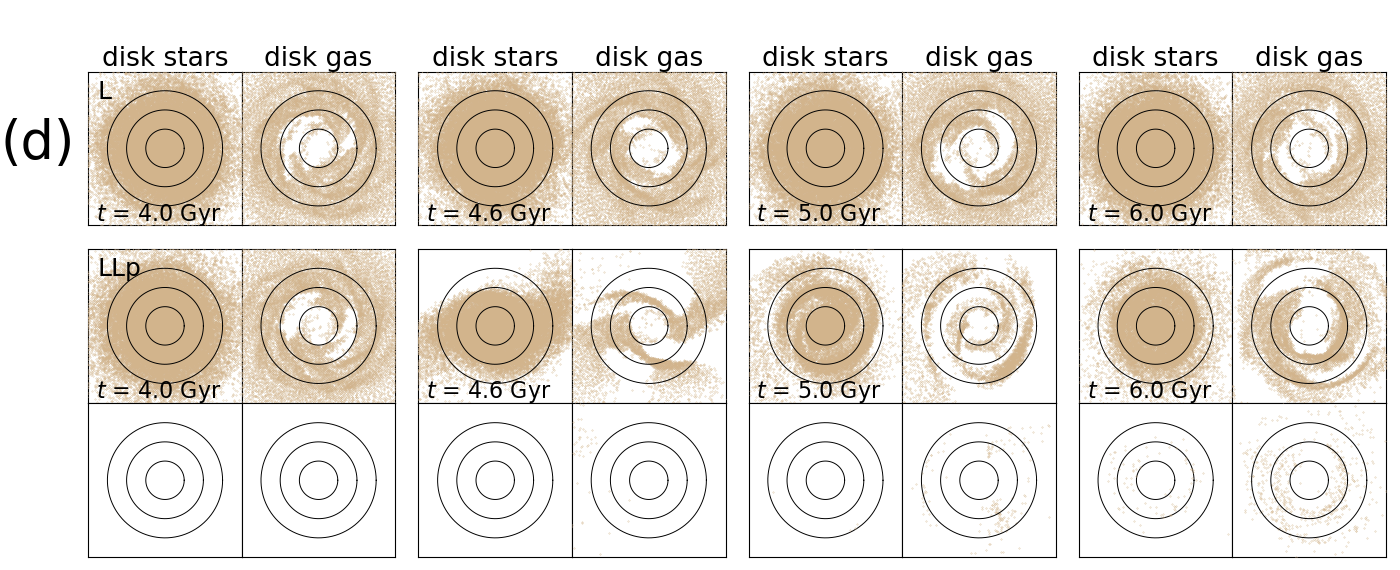}
\;\;
\includegraphics
[width=8.5cm]
{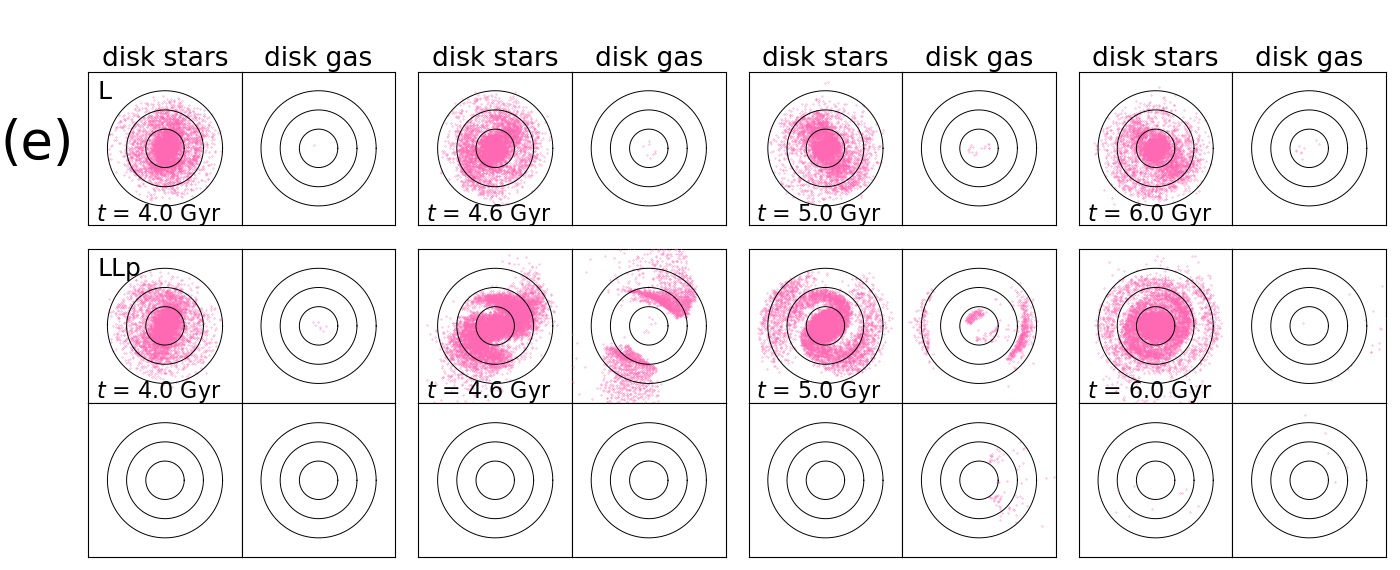}
\\
\vspace{0.3cm}
%---------------------
\includegraphics
[width=8.5cm]
{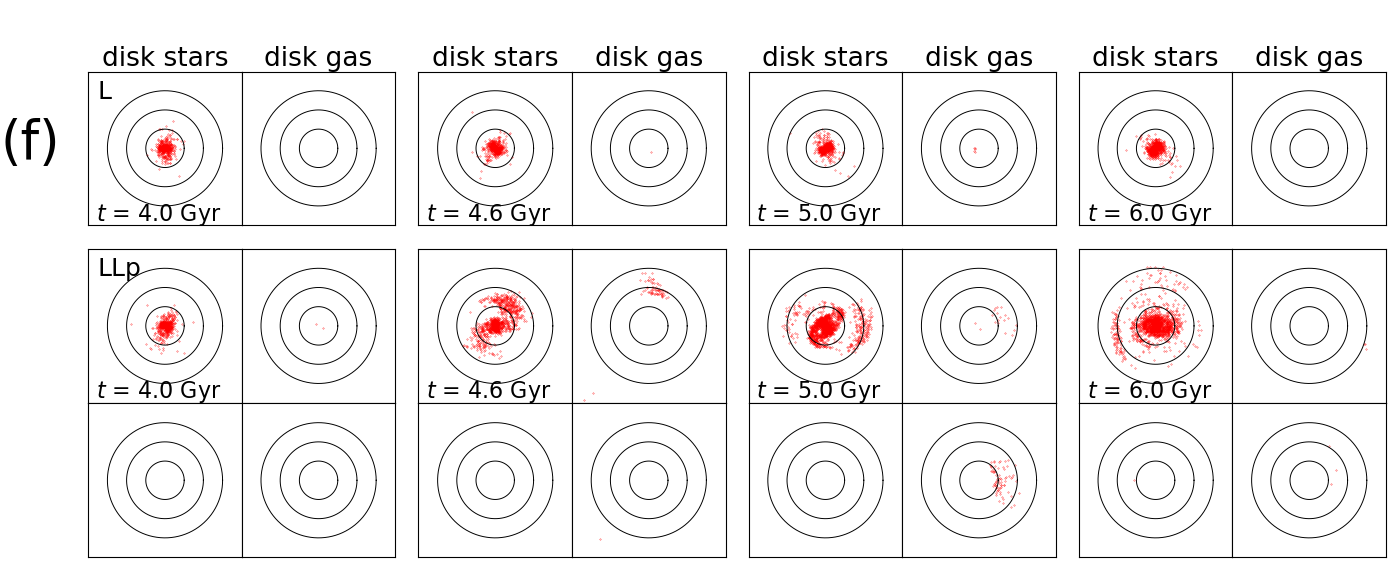}
\;\; 
\includegraphics
[width=8.5cm] 
{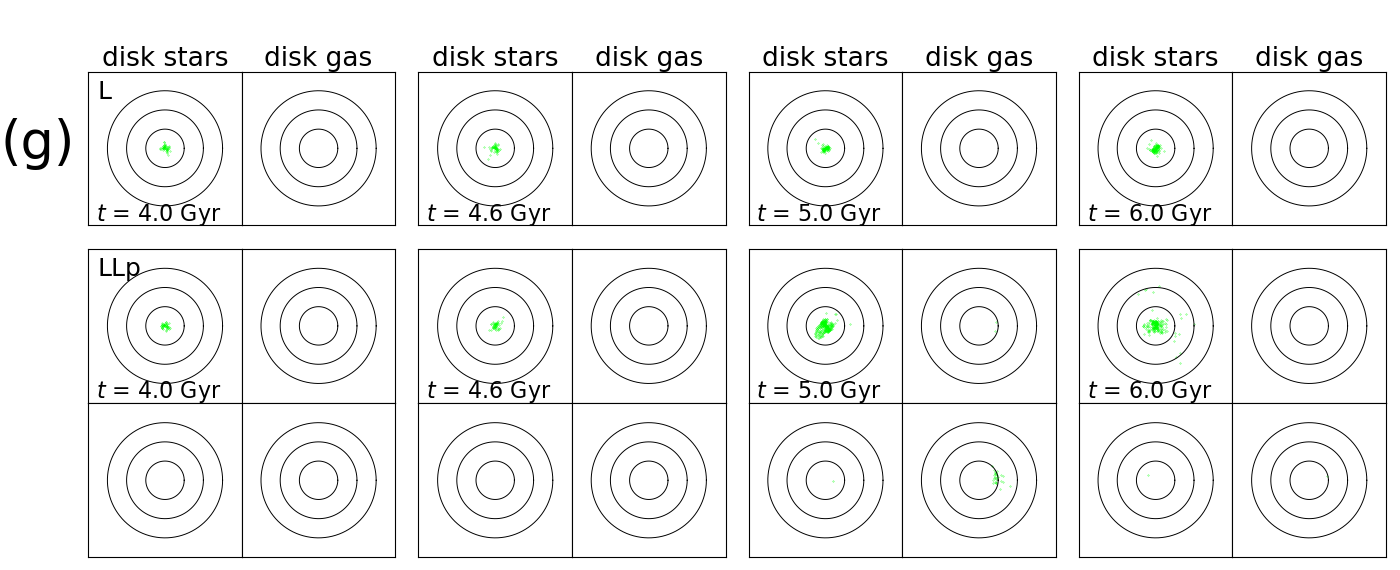}
\\
\caption{
(a) Disk star and gas particles of the target galaxy 
from runs~L and LLp    
at $t$ = 4, 4.6, 5, and 6~Gyr (from leftmost two columns 
to rightmost two columns in order, respectively). 
The disk particles of the target galaxy 
from run~L is shown in the top row; 
for run~LLp, the disk particles originally set as the target   
galaxy and those initially set as the companion galaxy 
but transferred to the target galaxy are displayed separately  
in the middle and bottom rows, respectively.
The particles are differently colored by the values 
of their circular velocities $v_{\rm c}$ 
as indicated in the color bar. 
(For simplicity, the particles having negative 
values of $v_{\rm c}$ are not distinguished,  
as they are not significant in runs~L and LLp, 
for the purpose of this figure.)    
The particles having the lowest values of $v_{\rm c}$ 
are plotted first and then those having higher values 
of $v_{\rm c}$ are overlaid on top in order.  
Each panel measures 40~kpc on a side. 
The concentric solid circles in each panel  
measure $R$ = 5, 10, and 15~kpc 
from the center of the target galaxy.
(b)-(g) Same as (a) 
but displaying the particles separately 
depending on the values of their circular velocities: 
0                    $\leq v_{\rm c} <$ 150 $\rm{km}~s^{-1}$ (b), 
150 $\rm{km}~s^{-1}$ $\leq v_{\rm c} <$ 200 $\rm{km}~s^{-1}$ (c),
200 $\rm{km}~s^{-1}$ $\leq v_{\rm c} <$ 250 $\rm{km}~s^{-1}$ (d),
250 $\rm{km}~s^{-1}$ $\leq v_{\rm c} <$ 300 $\rm{km}~s^{-1}$ (e),
300 $\rm{km}~s^{-1}$ $\leq v_{\rm c} <$ 350 $\rm{km}~s^{-1}$ (f), and 
350 $\rm{km}~s^{-1}$ $\leq v_{\rm c}$ (g).
\\
} 
\label{fig:fig10} 
\end{figure*}

%----------<< Figure 11: vc & vc+ >>----------
 
\begin{figure*} [!hbt]
\centering\includegraphics
[width=17.0cm]
{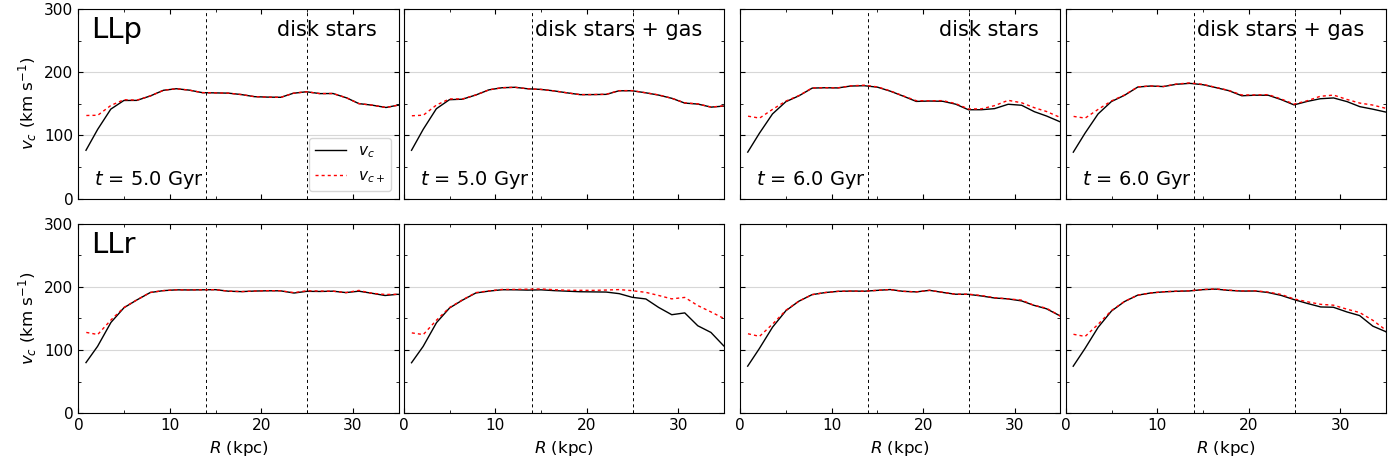}
\caption{
Comparison of the profiles of $v_{\rm{c}}$ and $v_{\rm{c+}}$ 
(black solid and red dotted curves, respectively) 
for the disk particles of the target galaxy 
from runs~LLp (top row) and LLr (bottom row).   
The left two columns show the profiles of the disk star particles  
and of the all disk particles, respectively, 
at $t = $~5~Gyr; the right two columns 
shows those at $t = $~6~Gyr.       
In each panel, the two vertical dotted lines  
mark $R_{14}$ and $R_{25}$. 
}
\label{fig:fig11} 
\end{figure*}

\subsection{Kinematics}

Here we discuss the kinematics of the  
disk star and gas particles of  
the target galaxy from our five runs. 
First we examine the time change of the 
circular and radial velocities of the disk particles  
in the following four subsections.  
Then we look into the evolution of the spin angular momentum 
of the disk material in the last subsection.

\subsubsection{Run~L}

Figures~\ref{fig:fig07}(a) and \ref{fig:fig08} (top row) 
present the profiles of the circular and radial  
velocities ($v_{\rm{c}}$ and $v_{\rm{R}}$, respectively)  
of the disk particles 
of the target galaxy while evolving in isolation.  
As seen in the figures, the profiles of the stars 
and gas deviate somewhat from the initial 
ones via secular evolution (see Figure~\ref{fig:fig01} for the 
initial profiles) - i.e., as a bar and spiral arms 
develop on the disk and star formation with 
the associated feedback comes into play.   
Although some changes appear in the velocity profiles 
as time passes, more for gas than for stars, 
in the relatively inner region of the disk  
(where star and structure formation occur more actively),     
the deviations are generally not significant 
throughout the run. 
(In general, the circular velocity profile of young disk stars 
could be different from that of (old) disk stars. 
For example, when they have just been formed, 
the profile may follow that of gas rather than that of stars. 
We do not display the profile of young stars separately 
in the figure, because their contribution to 
the overall (mass weighted) stellar velocity profile    
is not significant in our runs.  
It should also be noted that their contribution 
in luminosity could be more significant than in mass.)

In Figure~\ref{fig:fig09}, we compare the 
circular velocity profile of the disk stars 
and that of the entire disk material (stars + gas) 
at $t = $~5 and 6~Gyr (black solid curves).   
They appear to be quite similar to each other as expected,  
since stars in the target galaxy make up a large portion 
of the mass of the total disk.

Examining the values of $v_{\rm c}$ of the individual  
disk particles (top rows of (a)$-$(g) in 
Figure~\ref{fig:fig10}), 
stellar particles having relatively 
high values of $v_{\rm c}$ (out of the six bins 
as indicated by the color bar)    
appear dominantly in the inner region of the disk 
around the bar and the arms 
where the density of stars is high as well.  
Whereas, stars having lower values of $v_{\rm c}$ 
are spread more widely over the disk. 
It should be noted that many stars whose values of 
$v_{\rm c}$ falling in different bins overlap spatially. 
(E.g., stars lying in the innermost region of the disk 
have diverse values of $v_{\rm c}$, 
ranging from all of the six bins.)  
On the other hand, gas particles 
having lower values of $v_{\rm c}$ 
are found in the inner region of the disk 
and those having different values of $v_{\rm c}$ 
hardly overlap spatially.  
This is because, unlike stars, 
dynamically cold gas responds very quickly 
to the potential perturbations arisen by 
the galactic structures such as bar and spiral arm 
(\citealt{Roberts1969}; \citealt{Oh+2008}).  
Hence shocks can easily occur on the gas disk  
causing the gas to lose angular 
momentum (\citealt{Kim_Kim2014}; \citealt{Kim+2014}). 
For instance, the shocks driven by the asymmetric bar-potential 
can generate torques through which gas loses the 
angular momentum and moves radially inward;  
the spiral shocks can also work to remove the angular momentum 
of the gas facing with the ``pile up" of material.

\subsubsection{Run~LLp}

Figures~\ref{fig:fig07}(b) and \ref{fig:fig08} (middle row) show  
the profiles of $v_{\rm{c}}$ and $v_{\rm{R}}$, respectively, 
of the disk particles of the target galaxy experiencing 
a prograde collision with a late-type companion.  
First, examining the profiles of the disk stars 
in the region within $R_{25}$\footnote{We consider/compare 
the profile of $v_{\rm{c}}$ from each of our runs 
only in the region within $R_{25}$ throughout 
this paper - i.e., up to half the closest approach distance 
in the galaxy-galaxy encounter runs. 
That is because the outer part of the disk, 
further than $R_{25}$,   
of the target galaxy in some of our runs 
(particularly in runs~LLp and LEHp) is severely 
disturbed through the interaction.}  
(half the closest approach distance), 
there is a clear tendency for stars to lose $v_{\rm c}$ 
but gain both $v_{\rm{R+}}$ and $v_{\rm{R-}}$  
after the prograde collision.   
The decrease in circular motions for stars    
is mainly due to the deflection of 
the orbits by tidal disruption:  
(Although there are some particles transferring 
from the companion to the target,             
we have found that their contribution is not significant  
in shaping the overall velocity profiles.) 
As shown in Figure~\ref{fig:fig10} 
(bottom two rows of (a)$-$(g)), when the stellar disk 
of the target galaxy is pulled out by the tidal force,    
producing a bridge on the near-side (to the companion) 
and a counter-tail on the far-side,  
stars located in the outermost part of 
the disk are pulled first radially outward  
and then those in the next outermost part 
follow, while the disk rotates and moves. 
This may have caused the phase coherence of the 
stellar orbits along the arms, making the distribution 
of $v_{\rm c}$ as appeared in the figure. 
(Note that many stars with different values of $v_{\rm c}$ 
overlap spatially, as in run~L.)  
Since the stellar disk is strongly disturbed by the prograde 
collision and generate a substantial bridge and tail 
lasting a couple of giga years, 
the tendency of losing circular motion and 
gaining both radially inward and outward motions for stars 
(compared with those of before or without the collision) 
is maintained until $t = 6$~Gyr.

Next, as for the disk gas particles of the target galaxy,  
the circular and radial velocity profiles appear to vary 
more complicatedly after the collision,  
compared with those of stars:    
First of all, the amplitude of the circular velocity profile 
of gas does not generally decrease after the collision,  
but it rather fluctuates, sometimes making some inner peak(s)  
- e. g., as at $t = 5$~Gyr near $R = 3$~kpc 
(which will be discussed below). 
In addition, the amplitudes of the profiles of 
both $v_{\rm{R+}}$ and $v_{\rm{R-}}$ increase at $t = 5$~Gyr,   
until not long after the collision, 
and then slow down at $t = 6$~Gyr.  
During the collision, as shown in Figure~\ref{fig:fig10} 
(see also Figure~\ref{fig:fig06}(a)), 
the gas disk of the target galaxy is pulled 
out by the gravitational tide, 
starting from the outskirt of the disk.   
Besides, the gas disk is also shock-compressed 
by the collision with the gas disk of the companion, 
starting from the leading side of the disk, 
while it rotates and moves. 
As stated previously, since dynamically cold gas responds 
very quickly to the density perturbations generated on 
the disk, gas particles having different values 
of $v_{\rm c}$ hardly overlap spatially, 
in contrast with the case of stars.     
Locally, some disk gas particles can gain the 
circular velocities through the collision,  
under the influence of the collisional compression and 
the shocks generated on the disk in addition to that of gravity.  
For example, the gas particles of the target galaxy 
at $t = 5$~Gyr lying along the asymmetric inner arm around $R =$ 3~kpc 
(those shown in pink in the middle row sixth panel 
from the left in Figure~\ref{fig:fig10}(a)) 
gain circular velocities, resulting in the peak around 
the radius in the profile of $v_{\rm c}$. 
Interestingly, at $t = 5$~Gyr, 
the gas particles transferring to the target galaxy 
through the bridge (formed out of the 
gas disk of the companion) reaching the region near 
$R = $ 5~kpc (measuring from the center of the target; 
those shown in green and red in the bottom row sixth panel 
from the left in Figure~\ref{fig:fig10}(a)) have 
particularly high circular velocities.  
It has occurred because the direction of progress 
of the gas bridge happens to be aligned with the direction 
of the local circular velocity in the target galaxy. 
In this particular case, however, the amount of the 
transferring gas is revealed to be not enough 
to make a significant contribution 
to the profile of $v_{\rm c}$.

We compare in Figure~\ref{fig:fig09} (top row) 
the circular velocity profiles of the disk star particles   
and of the all disk particles of the target galaxy
at $t = $~5 and 6~Gyr (blue dashed curves). 
The profile of the all disk particles generally follows 
that of the stars only, as in run~L.  
It is because stars are dominant in mass on the disk and 
the encounter is not so strong as to make the system 
to be out of equilibrium.    
It is also clearly shown that both profiles of the stars and 
the star + gas from this run are well below those from run~L 
after the collision, except in the central region. 
In addition, we examine how the circular velocity profile 
changes when excluding counter-rotating disk particles. 
Figure~\ref{fig:fig11} displays that 
the profiles of $v_{\rm{c+}}$ of the stars   
and of the stars + gas (excluding counter-rotating ones) are 
quite similar to the profiles of $v_{\rm{c}}$ 
after the prograde collision, except at the central region. 
(Note that some difference between the profiles 
of $v_{\rm{c}}$ and $v_{\rm{c+}}$ also appears 
at the central region of the isolated target galaxy, 
as similarly as those in run~LLr (bottom row)).

\subsubsection{Run~LLr}

When the target galaxy experiences a retrograde collision 
with a late-type companion,   
the profile of $v_{\rm c}$ (Figure~\ref{fig:fig07}(c))   
of the disk stars within $R_{25}$ 
remains relatively unchanged after the collision.  
The amplitudes of both profiles of $v_{\rm{R+}}$ and 
$v_{\rm{R-}}$ of the stars (bottom row of Figure~\ref{fig:fig08}) 
also do not change much through the retrograde collision.  
As for the disk gas within $R_{25}$, 
the profiles of $v_{\rm{c}}$, $v_{\rm{R+}}$, and $v_{\rm{R-}}$  
change somewhat after the collision (except during 
or shortly after the collision), but not as greatly as in run~LLp. 
Figure~\ref{fig:fig09} (top row) also shows that 
the circular velocity profiles of the stars and 
of the star + gas from this run do not differ much 
from those from run~L at $t =$~5 and 6~Gyr.  
Figure~\ref{fig:fig11} (bottom row) presents the profiles 
of $v_{\rm{c}}$ and $v_{\rm{c+}}$ of the disk particles. 
The two profiles within $R_{25}$ are generally 
similar, except at the central region of the disk.

This result indicates that the retrograde collision 
considered here has not been able to drive significant 
dynamical response in the target galaxy, 
in contrast with the case of the prograde collision, 
despite of the equally close encounter. 
Therefore, it can be inferred that 
when a pair of galaxies are observed to be close to each other,  
the effects on the kinematics of a system 
by the interaction with its neighbour   
could be too weak to be measured,  
depending on the geometry of the encounter.

\subsubsection{Runs~LEHp and LEHr}

Let's first consider run~LEHp, where  
the target galaxy interacts with an 
early-type companion in prograde sense. 
As shown in Figures~\ref{fig:fig07}(d) and \ref{fig:fig09}, 
the profile of $v_{\rm c}$ of the disk stars of the target galaxy 
evolves somewhat similarly to that from run~LLp,  
after the collision   
- i. e., stars generally lose $v_{\rm c}$ after the collision. 
Since the gravitational pull exerted upon the 
target galaxy depends on the mass of the 
companion, and which is the same in runs~LEHp and LLp,  
the stellar orbits in the target galaxy 
from both runs are influenced 
in a similar way through the prograde collision.

For the disk gas of the target galaxy from run~LEHp, 
the profile of $v_{\rm c}$ becomes somewhat bumpy, 
without a marked decrease, after the collision.     
This trend has also been found in that of the gas from run LLp. 
However, the specific shape of the profile from this run     
evolves differently from that from run~LLp after the collision. 
This may have been caused by the different 
hydrodynamic interactions in these runs.

Finally for run~LEHr 
(Figures~\ref{fig:fig07}(e) and \ref{fig:fig09}),   
where the target galaxy experiences a 
retrograde collision with an early-type companion, 
the profile of $v_{\rm c}$  
of the disk stars of the target galaxy 
remains almost unchanged after the collision,  
similarly as that from run~LLr.   
The velocity profile of the gas 
varies after the collision but only slightly. 
As pointed out in Section~3.2.3, 
it means that the dynamics of the target galaxy 
in this run, as well as in run~LLr, 
is not affected significantly by the retrograde collision, 
except the outskirts of the disk.

%----------<< Figure 12 : Spin Angular Momentum >>----------
 
\begin{figure*} [!hbt]
\centering\includegraphics
[width=14cm]
{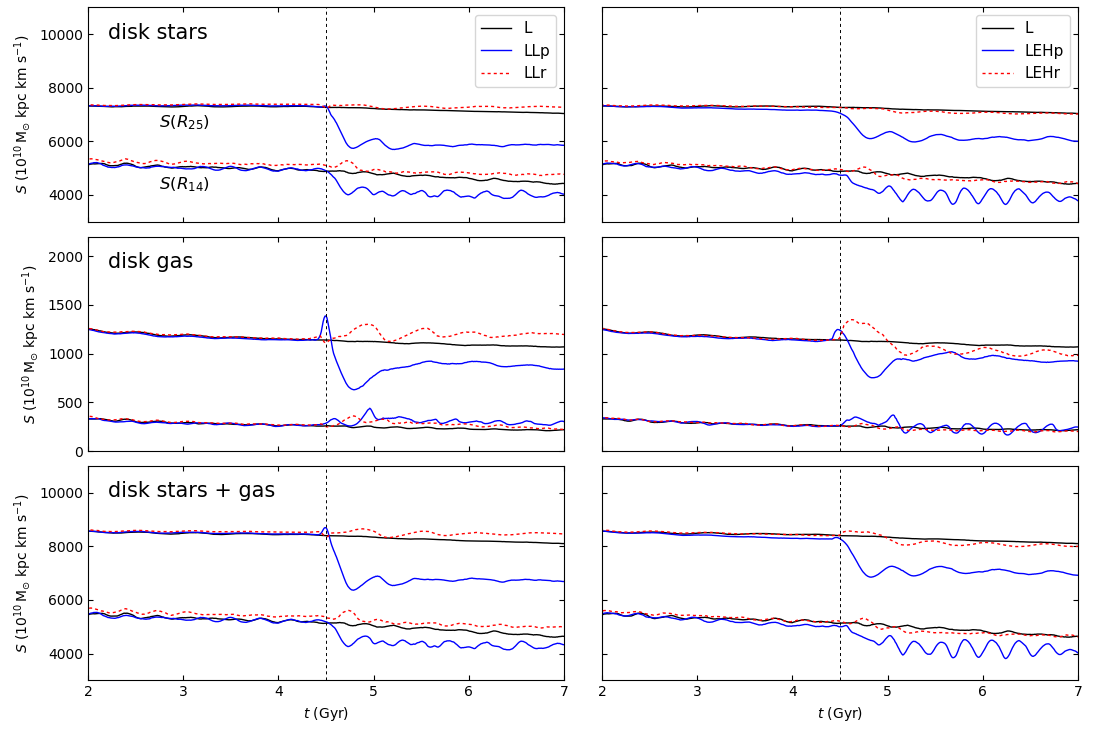}
\caption{
Spin angular momentum of the disk 
of the target galaxy with respect to time 
from runs~L, LLp and LLr (left column) 
and runs~L, LEHp, and LEHr (right column). 
The top through bottom rows show  
the angular momentum of the disk, considering 
the disk stars (top row), gas (middle row), 
and stars $+$ gas (bottom row) particles  
enclosed in a cylindrical region of a 
radius $R$ ($S(R)$, with $|z| \leq$ 20~kpc). 
In each panel, the upper and lower three curves 
display $S(R_{25})$ and $S(R_{14})$, respectively.
(The particles captured by the target 
from the companion are included, if there are any.)    
The vertical dotted line indicates the closest approach
time, $t = $ 4.5~Gyr, between the target and the companion 
galaxies in the galaxy-galaxy encounter runs.  
}
\label{fig:fig12} 
\end{figure*}

%----------<< Figure 13 : Specific Spin Angular Momentum >>----------
 
\begin{figure*} [!hbt]
\centering\includegraphics 
[width=14cm]
{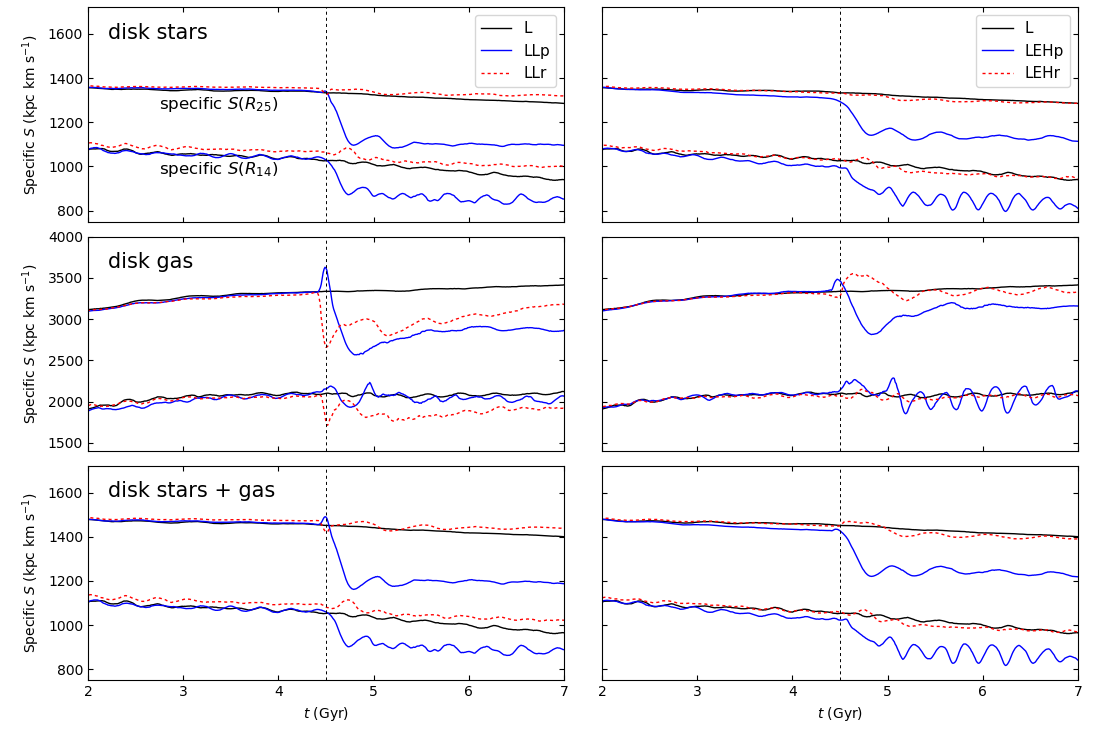}
\caption{
Same as Figure~\ref{fig:fig12}, 
but for the specific angular momentum of the 
disk of the target galaxy. 
}
\label{fig:fig13} 
\end{figure*}

\subsubsection{Angular Momentum}

We calculate the spin angular momentum $S$ of 
the target galaxy disk from all of our runs.  
For the estimation of $S$($R$),   
we consider the disk particles of the target galaxy 
enclosed within a cylindrical region of a radius $R$ 
(with $|z| \leq$ 20~kpc).
We present in Figure~\ref{fig:fig12} 
the evolutions of $S$($R_{25}$) and $S$($R_{14}$).  
It is shown that $S$ of the star disk and of the star $+$ gas disk 
of the target galaxy decrease drastically 
after the prograde collision (runs~LLp and LEHp),  
but not after the retrograde collision (runs~LLr and LEHr). 
Specifically, the value of $S$($R_{25}$) of the star $+$ gas disk 
at $t = $ 5.5~Gyr (i.e., 1~Gyr after the collision) 
is about 20~\%(15~\%) smaller in run~LLp(LEHp) 
than that at $t = $ 3.5~Gyr (1~Gyr before the collision).  
In Figure~\ref{fig:fig13}, we also present 
the specific angular momentum of the disk of the target galaxy, 
by dividing $S$ by the total mass of the enclosed particles in each case. 
The specific angular momentum of the disk also decreases clearly  
after the prograde collision, similarly to the angular momentum. 
This phenomenon can be interpreted as follows:
Initially, in the center-of-mass (CM) frame, 
two galaxies are far apart and approach linearly each other. 
After the encounter, those galaxies rotate about the CM
with the orbital angular momentum $L$, 
and the total angular momentum of this system is given by 
the vector sum of $\vec{S}$ and $\vec{L}$ of those galaxies. 
The sum is additive in magnitude in the prograde case, 
but subtractive in the retrograde case. 
This results in the decrease of $S$ of the 
target galaxy disk in the prograde case 
while the increase of that in the retrograde case. 
This justifies our result showing the different behavior of $S$ 
in the prograde case against the retrograde case. 
The contribution of the tidal effects on the target galaxy disk 
discussed in the previous sections decreases $S$ further, 
in such a way that the decrease of $S$ in the prograde case is enhanced 
and the increase of $S$ in the retrograde case is reduced.
The change of $L$ is biggest right after the encounter, and
one can see such a drastic change of $S$ as 
shown in Figure~\ref{fig:fig12}. 
There could be also the tidal effects due to the  
bulge and the DM halo reducing $S$ of the galaxy disks 
in both of the prograde and retrograde cases,
but those contributions are not significant.  
We will discuss such subdominant effects briefly in 
Appendix B.

%====================================
%            Sec. 4
%==================================== 

\section{Summary and Discussion}

Influenced by observational studies 
that found some correlations between the 
spin magnitudes in pairs of galaxies 
(\citealt{Sodi+2010}; \citealt{Lee+2018}),  
we have carried out this numerical study     
in order to examine how the spin of LTGs 
is affected by interactions, in a more direct way.      
As the first in a series of (planned) work,  
we have constructed four simulations of co-planner encounters  
between a late-type target galaxy (model~L) and 
an equally massive companion galaxy, whose 
morphology is either a late- or an early-type 
(model~L or EH, respectively), 
in prograde or retrograde sense. 
They are the LLp, LLr, LEHp, and LEHr runs.  
We have also set up one simulation of the evolution 
of the target galaxy in isolation, run~L, for comparison.

By analysing the circular velocities of the disk particles 
of the target galaxy (up to $R_{25}$) over time 
from each of our five runs, 
we have obtained the following results:

1. Disk stars of the late-type target galaxy 
tend to lose the circular velocities 
(and to gain both positive and 
negative radial velocities) 
through the prograde collision in runs~LLp and LEHp, 
regardless of the type of the companion. 
It is mainly due to the deflection of 
the orbits caused by the tidal disruption.    
Since the masses of the late- and the early-type companions 
(having the identical DM halo component) are the same,     
the individual disk stars (which are mainly affected 
by gravitational force) of the target galaxy in these runs 
dynamically evolve in a similar way through the collision.

2. Disk gas of the target galaxy 
dose not generally lose the circular velocity through 
the prograde collision in runs~LLp and LEHp.   
The circular velocity of the gas vary 
in a more complicated fashion, than those of the stars,    
due to the combined effects of the gravitational 
and the hydrodynamic interactions.

3. In case of the retrograde collision (runs~LLr and LEHr),  
the circular velocities of the disk stars of the target galaxy 
remain almost unchanged after the collision,  
as similarly as those in the isolated target galaxy (run~L). 
The circular velocity of the gas changes 
after the retrograde collision, 
but not as greatly as in runs~LLp and LEHp, 
except during/near the collision.      
It indicates that the retrograde collision in these runs 
does not drive a significant dynamical 
response in the target galaxy, 
irrespective of the type of the companion.

Since stars in our target galaxy make up a large portion of 
the mass of the total disk, as does in many well observed LTGs,    
the circular velocity profile of the entire disk  
material (stars + gas) generally follows that of the stars 
in all of our runs. 
It is also because the encounters considered here are not 
so strong as to make the systems to be out of equilibrium.    
Taken together, our simulation results indicate that 
the overall spin of a LTG can decrease through a prograde collision,  
but hardly through a retrograde collision, 
regradless of the morphology of the companion. 
It was also demonstrated in the time change of the 
spin angular momentum of the disk of the target galaxy. 
We have found that the angular momentum 
clearly decreases after the prograde collision, 
but not after the retrograde collision. 
This is due to the combined contributions 
of the tidal effect on the target disk
and the spin-orbits interactions 
(followed by the total angular momentum conservation)
of the two galaxies. 
Specifically, the spin angular momentum (within $R_{25}$) 
decreased by $\sim$20\% and $\sim$15\% after the 
prograde collision in runs~LLp and LEHp, respectively. 
We can thus infer that, for a sample of LTGs 
which are observed to have a close neighbour of a comparable mass,  
the tendency of decrease of the spin of the LTGs 
(as the separation distance decreases) would be more 
clearly measured when the sample includes many LTGs that 
experiences and/or have experienced 
one or more prograde collisions.

The results from this work may not be 
generalized to different cases.  
For example, in case where a late-type target galaxy 
interacts with a companion galaxy of a highly unequal mass, 
a significant net amount of material can transfer 
from one galaxy to the other. 
Let us recall that, in run~LLp at $t =$ 5~Gyr, 
there were some gas particles transferring 
from the companion to the target through the narrow bridge,   
having much higher circular velocities 
than surrounding particles.  
Their higher circular velocities could be made 
because the direction of progress of the bridge happened to be 
aligned with the direction of the local circular velocity. 
The amount of the transferring material in run~LLp   
was not enough to make a meaningful impact in shaping  
the circular velocity profile of the entire disk material.         
However, in different situations 
such as interactions between unequal mass galaxies,  
a sufficient amount of material might be able to transfer 
to the target galaxy with high circular velocities, 
resulting in a peak in the circular velocity profile 
at the corresponding radius.

Another example, when a gas-rich late-type target galaxy     
interacts strongly with another gas-rich galaxy, 
hydrodynamic effects of the interaction 
may play a much more influential role than in our runs. 
In this case, the amplitude of the circular velocity profile 
of the overall disk material of the target galaxy    
might not simply decrease after the collision, 
unlike in our runs~LLp and LEHp (where tidal disruption 
plays a bigger role to decrease the amplitude), 
but changes in a very complex fashion depending on the 
particular circumstance of each case.

We end this work by emphasizing the significance of 
interactions with neighbouring galaxies 
in determining   
the spin of LTGs at current stage,   
as demonstrated by our runs~LLp and LEHp.  
Since most galaxies experience multiple interactions 
over the course of their lifetime, 
the accumulated effects of galaxy-galaxy interactions  
at later stages after the acquisition of the 
internal angular momentum     
will play a crucial role in the evolution of 
the total angular momentum of galaxies.  
We will continue studying this subject 
in our future work, 
considering more diverse interactions 
including the cases stated above 
and also major and minor 
mergers for a more in-depth understanding of 
the evolution of galactic spin 
in various environments. 
\\

%====================================
%            Acknowledgement
%====================================

We thank the referee Santi Roca-F{\`a}brega and 
the anonymous referee very much  
for insightful comments and helpful suggestions  
which improved this paper significantly.  
J.-S. H. is grateful to Juhan Kim and Yonghwi Kim 
for helpful discussions. 
We appreciate Joshua E. Barnes making the ZENO code available
and Volker Springel for providing us with GADGET-3.  
We thank the Korea Institute for Advanced Study for providing
computing resources (Higgs Server and 
KIAS Center for Advanced Computation
Linux Cluster System) for this study.
This work was supported by 
Basic Science Research Program 
through the National Research Foundation of Korea (NRF)  
funded by the Ministry of Education under the Grant Nos. 
NRF-2018R1D1A1B07048156 (J.-S. Hwang) and  
NRF-2018R1D1A1B07050701 (J.-S. Hwang), and 
NRF-2020R1I1A1A01072816 (S.-h. Nam), 
and also funded by the Ministry of Science and ICT 
under the Grant No. 2020R1A2C3009918 (S.-h. Nam).

%----------<< Figure 14 : LLr2 large-scale ss >>----------
 
\begin{figure*} [!hbt]
\centering\includegraphics 
[width=15cm]
{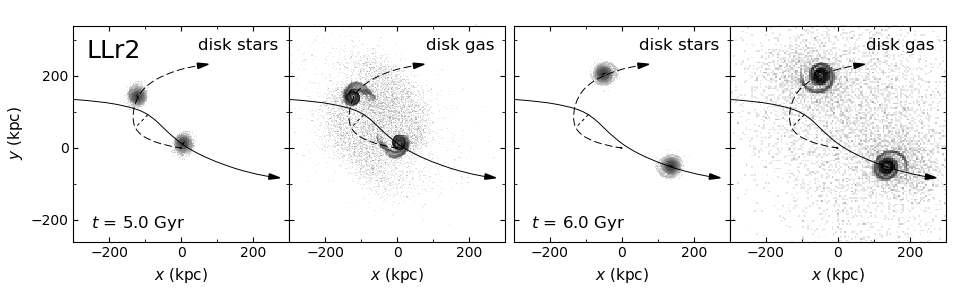}
\caption{
Same as Figure~\ref{fig:fig05}, but for run~LLr2.
}
\label{fig:fig14} 
\end{figure*}

%----------<< Figure 15 : LLr2 ss >>----------

\begin{figure*} [!hbt]
\centering
\includegraphics
[width=18.0cm]
{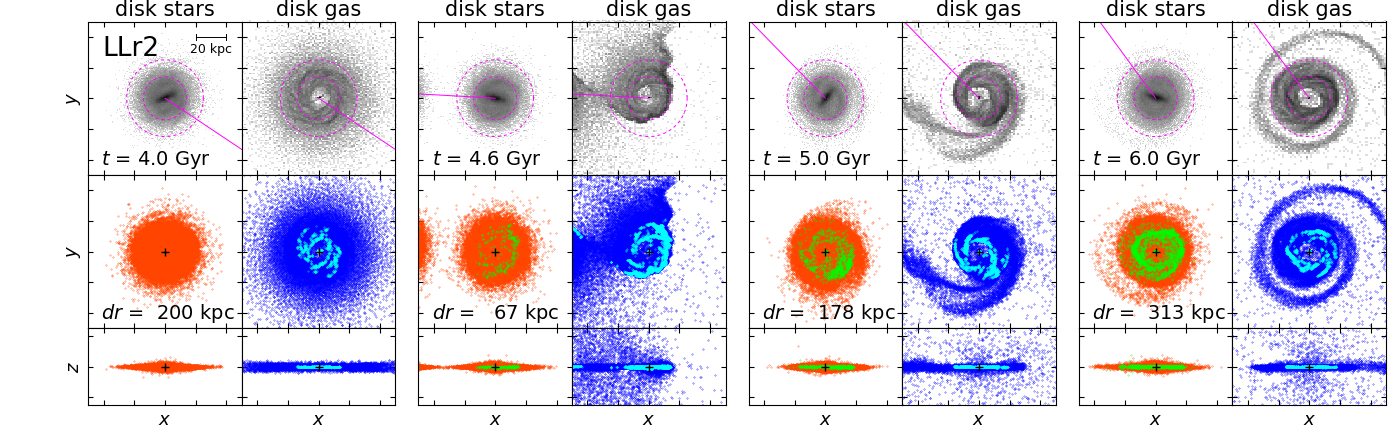}
\caption{
Same as Figure~\ref{fig:fig03}, but for run~LLr2.
}
\label{fig:fig15} 
\end{figure*}

%----------<< Figure 16:  LLr2 Vc: stars -- stars+gas >>----------
 
\begin{figure*} [!hbt]
\centering\includegraphics
[width=17.0cm]
{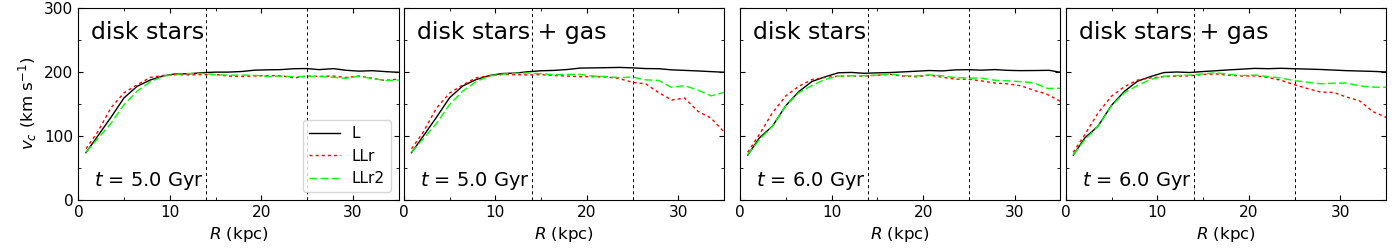}
\caption{
Same as Figure~\ref{fig:fig09}, but for runs~L, LLr, and LLr2.
}
\label{fig:fig16} 
\end{figure*}

%----------<< Figure 17 : Spin Angular Momentum - Bulge & Halo >>----------
 
\begin{figure*} [!hbt]
\centering\includegraphics
[width=14cm]
{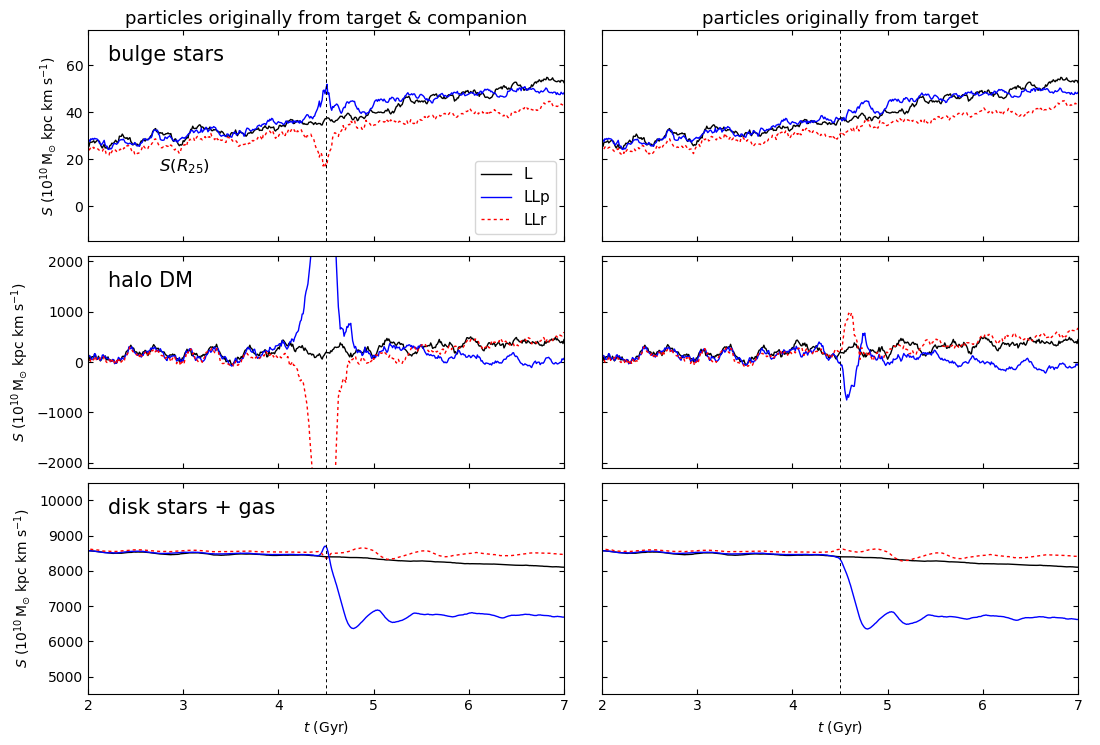}
\caption{
Spin angular momentum $S(R_{25})$ of the stellar bulge (top row), 
the DM halo (middle row), and the disk (bottom row)   
of the target galaxy with respect to time 
from runs~L, LLp and LLr. 
In the calculation of $S(R_{25})$, 
only the mass within $R_{25}$ and $|z| \leq$~20~kpc  
from the center of the target is considered. 
The left and right columns show the change of $S(R_{25})$, 
when all particles are used  
or when only the particles originally belonging to the target 
galaxy are used, respectively. 
In each panel the vertical dotted line indicates 
the closed approach time, $t$ = 4.5~Gyr, between 
the target and the companion galaxies in the encounter runs. 
}
\label{fig:fig17} 
\end{figure*}

%====================================
%            Appendix A
%====================================

\begin{center}
\MakeUppercase{{Appendix A}\\}
{Run~LLr2}\\
\end{center}

We have performed one more simulation, ``run~LLr2". 
The only difference between the initial conditions of this run 
and run~LLr is the direction of the initial spin 
of the companion galaxy - i. e.,  
it is set with a counterclockwise direction in run~LLr2.  
Therefore, when the target and the companion meet most closely, 
both galaxies experience the encounter with the other   
as retrograde.

We present the snapshots (Figures~\ref{fig:fig14}, 
and \ref{fig:fig15}) and 
the circular velocity profiles (Figure~\ref{fig:fig16}) from run~LLr2. 
Both target and the companion galaxies 
do not develop substantial bridges and tails 
after the retrograde encounter as expected. 
The overall circular velocity of the disk material  
of the target galaxy does not decrease much 
after the collision as shown in Figure~\ref{fig:fig16}.  
The total spin angular momentum becomes doubles in this case so that 
the change of the spin angular momentum is milder than that of run LLr 
(where the spin sum is subtractive due to the opposite spin directions) 
as one can see from the figure.

%====================================
%            Appendix B
%====================================

\begin{center}
\MakeUppercase{Appendix B\\}
{Angular moment of various components}\\
\end{center}

In Figure~\ref{fig:fig17} we present    
the spin angular momentum $S(R_{25})$  
of the various components of the target galaxy 
from run~L, LLp, and LLr. 
The figure indicates that most of the spin angular momentum 
of the target galaxy's disk is not transferred to 
different components of the target galaxy; 
it is likely to be transferred to the orbital angular momentum 
of the two interacting galaxies. 
In order to have full access to the mechanism of 
angular momentum transfer of interacting galaxies, 
more extensive analysis are need to be done. 
\\

%====================================
%            Refferences
%====================================

\end{document}